\documentclass[aps,showpacs,superscriptaddress,tightenlines,eqsecnum,nofootinbib]{revtex4}
\usepackage{graphicx,amssymb,amsbsy,bm,amsmath}
\usepackage{dsfont}
\usepackage{hyperref}

\newcommand{\vx}{\ensuremath{\vec{x}}}
\newcommand{\vk}{\ensuremath{\vec{k}}}

\newcommand{\ovom}{\overline{\omega}(k)}

\newcommand{\be}{\begin{equation}}
\newcommand{\ee}{\end{equation}}
\newcommand{\bea}{\begin{eqnarray}}
\newcommand{\eea}{\end{eqnarray}}

\begin{document}
\title{Non equilibrium dynamics of mixing, oscillations   and equilibration: \\
a model study. }
\author{D. Boyanovsky}
\email{boyan@pitt.edu} \affiliation{Department of Physics and
Astronomy, University of Pittsburgh, Pittsburgh, Pennsylvania 15260,
USA}
\author{C. M. Ho} \email{cmho@phyast.pitt.edu}
\affiliation{Department of Physics and Astronomy, University of
Pittsburgh, Pittsburgh, Pennsylvania 15260, USA}
\date{\today}

\begin{abstract}
The non-equilibrium dynamics of mixing, oscillations and
equilibration is studied in a  field theory of flavored neutral
mesons that effectively models  two flavors of  mixed neutrinos, in
interaction with other mesons that represent a thermal bath of
hadrons or quarks and charged leptons.  This model describes the
general features of neutrino mixing and
  relaxation via charged currents in a medium. The reduced density matrix
  and the
non-equilibrium effective action that describes the propagation of
neutrinos is obtained by integrating out the bath degrees of
freedom. We obtain the dispersion relations, mixing angles and
relaxation rates of ``neutrino'' quasiparticles. The dispersion
relations and mixing angles are of the same form as those of
neutrinos in the medium, and the relaxation rates are given by
$\Gamma_1(k)    = \Gamma_{ee}(k)
\cos^2\theta_m(k)+\Gamma_{\mu\mu}(k)\sin^2\theta_m(k)
 ~;~ \Gamma_2(k)   =     \Gamma_{\mu\mu}(k)
\cos^2\theta_m(k)+\Gamma_{ee}(k)\sin^2\theta_m(k) $ where
$\Gamma_{\alpha\alpha}(k)$ are the relaxation rates of the flavor
fields in \emph{absence} of mixing, and $\theta_m(k)$ is the mixing
angle in the medium.  A Weisskopf-Wigner approximation that
describes the asymptotic time evolution in terms of a non-hermitian
Hamiltonian is derived. At long time $>>\Gamma^{-1}_{1,2}$
``neutrinos''   equilibrate with the bath. The equilibrium density
matrix is nearly diagonal in the basis of eigenstates of an
\emph{effective   Hamiltonian that includes self-energy corrections
in the medium}. The equilibration of ``sterile neutrinos'' via
active-sterile mixing is discussed.
\end{abstract}


\maketitle

\section{Introduction}\label{sec:intro}

 Neutrinos   are   the central link between particle and nuclear physics,
astrophysics and cosmology\cite{book1,book2,book3,raffelt,kayserrev}
and the experimental confirmation  of neutrino mixing and
oscillations provide a first evidence for physics beyond the
Standard Model. Neutrino mixing provides an explanation for the
solar neutrino problem\cite{MSWI,haxton1,haxton,langacker}, plays a
fundamental role in the physics of core-collapse
supernovae\cite{wolf,haxfuller,panta,kotake,fullerast,garysn,burro,praka}
and in cosmology\cite{dolgovrev}: in big bang nucleosynthesis
(BBN)\cite{steigman}, baryogenesis through
leptogenesis\cite{fukugita,yanagida,buch,pila}, structure
formation\cite{pastor,dodelson1,dodelson2,dm,aba}   and possible
dark matter candidate\cite{dodelson2,dm,aba}.

The \emph{non-equilibrium dynamics} of neutrino mixing, oscillations
and equilibration is of  fundamental importance in all of these
  settings. Neutrinos are produced as ``flavor eigenstates''
in weak interaction vertices, but propagate as a linear
superposition of mass eigenstates. This is the origin of neutrino
oscillations. Weak interaction collisional processes are diagonal in
flavor leading to a competition between production, relaxation and
propagation which results in a complex and rich dynamics.

Beginning with pioneering work on neutrino mixing in
media\cite{dol,stodolsky,raf,mac,kari}, the study of the dynamical
evolution has been typically cast in terms of single particle
``flavor states'' or matrix of densities that involve either a
non-relativistic treatment of neutrinos or consider flavor neutrinos
as massless. The main result that follows from these studies is a
simplified set of Bloch equations with a semi-phenomenological
damping factor (for a thorough review see\cite{dolgovrev}).

Most of these approaches involve in some form the concept of
distribution functions for ``flavor states'', presumably these are
obtained as expectation values of Fock number operators associated
with flavor states. However, there are several conceptual
difficulties associated with flavor Fock states, still being
debated\cite{giunti1,giunti2,blasone,cardallkine,fujii,ji,li,chargedlepton}.

The importance of neutrino mixing and oscillations, relaxation and
equilibration in all of these timely aspects of cosmology and
astroparticle physics warrant a deeper scrutiny of the
non-equilibrium phenomena firmly based on quantum field theory.

{\bf The goals of this article:} Our ultimate goal is to study the
non-equilibrium dynamics of oscillation, relaxation and
equilibration directly in the quantum field theory of weak
interactions bypassing the ambiguities associated with the
definition of flavor Fock states. We seek to understand the nature
of the equilibrium state: the \emph{free field Hamiltonian} is
diagonal in the mass basis, but the interactions are diagonal in the
flavor basis, however, equilibration requires interactions, hence
there is a competition between mass and flavor basis, which leads to
the question of which is the basis in which the equilibrium density
matrix is diagonal. Another goal is to obtain the dispersion
relations and the \emph{relaxation rates} of the correct
quasiparticle excitations in the medium.

In this article we make progress towards these goals by studying   a
simpler model of two ``flavored'' \emph{mesons} representing the
electron and muon neutrinos that mix via an off-diagonal mass matrix
and interact with other \emph{mesons} which represent either hadrons
(neutrons and protons) or quarks and charged leptons via an
interaction vertex that models the charged current weak interaction.
The meson fields that model hadrons (or quarks) and charged leptons
are taken as a \emph{bath in thermal equilibrium}. In the standard
model the assumption that hadrons (or quarks) and charged leptons
can be considered as a bath in thermal equilibrium is warranted by
the fact that their strong and electromagnetic interactions
guarantee faster equilibration rates than those of neutrinos.

This model  bears the most relevant characteristics of the standard
model Lagrangian augmented by an off diagonal neutrino mass matrix
and will be seen to yield a remarkably faithful description of
oscillation and relaxational dynamics in a thermal medium at high
temperature. It effectively describes the thermalization dynamics of
neutrinos in a medium at high temperature such as the early Universe
for $T \gtrsim 3 \, \textrm{MeV}$\cite{dolgovrev,steigman,book3}.

Furthermore, Dolgov et. al.\cite{dolokun}  argue  that the spinor
nature of the neutrinos is not relevant to describe the dynamics of
mixing at high energies, thus we expect that this model captures the
relevant dynamics.


An exception is the case of neutrinos in supernovae, a situation in
which neutrino degeneracy, hence Pauli blocking,  becomes important
and requires a full treatment of the fermionic aspects of neutrinos.
Certainly the \emph{quantitative} aspects such as relaxation rates
must necessarily depend on the fermionic nature. However,  we expect
that a bosonic model will capture, or at minimum provide a guiding
example, of the most general aspects of the non-equilibrium
dynamics. The results found in our study lend support to this
expectation.


While meson mixing has been studied previously\cite{ji2}, mainly
motivated by mixing in  the neutral kaon and pseudoscalar
$\eta,\eta'$ systems,
  our focus is different in that we study the real time
dynamics of oscillation, relaxation and equilibration  in a
\emph{thermal medium} at high temperature including radiative
corrections with a long view towards understanding general aspects
that apply to neutrino physics in the high temperature environment
of the early Universe.

While neutrino equilibration in the early Universe for $T \gtrsim 3
\, \textrm{MeV}$ prior to BBN is
undisputable\cite{dolgovrev,steigman,book3}, the main questions that
we address in this article are whether the equilibrium density
matrix is diagonal in the flavor or mass basis and the relation
between the relaxation rates of the propagating modes in the medium.

\vspace{2mm}

{\bf The strategy:} the meson fields that model flavor neutrinos are
treated as the ``system'' while those that describe hadrons (or
quarks) and charged leptons, as the ``bath'' in thermal equilibrium.
An initial density matrix is evolved in time and the ``bath'' fields
are integrated out up to second order in the coupling to the system,
yielding a ``reduced density matrix'' which describes the dynamics
of correlation functions solely of system fields (neutrinos). This
program  pioneered by Feynman and Vernon\cite{feyver} for coupled
oscillators (see also\cite{leggett,boyalamo}) is carried out in the
interacting theory by  implementing the closed-time path-integral
representation of a time evolved density matrix\cite{schwinger}.
This method yields the \emph{real time non-equilibrium effective
action}\cite{hoboydavey}   including the   self-energy which yields
the ``index of refraction'' correction to the mixing angles and
dispersion relations\cite{notzold} in the medium  and the decay and
relaxation rates of the quasiparticle excitations. The
non-equilibrium effective action thus obtained yields the time
evolution of correlation and distribution functions and expectation
values in the reduced density matrix\cite{hoboydavey}. The approach
to equilibrium is determined by the long time behavior of the two
point correlation function and its equal time limit, the one-body
density matrix. The most general aspects of the dynamics of   mixing
and equilibration  are completely determined by the spectral
properties of the correlators of the bath degrees of freedom in
equilibrium.

\vspace{2mm}

{\bf Brief summary of results:}

\begin{itemize}
\item{ We discuss the ambiguities in the definition of
flavor Fock operators,   states and distribution functions.}

\item{The non-equilibrium effective action is obtained up to second order in the coupling $G \sim G_F$ between the ``system''
(neutrinos) and the bath (hadrons, quarks and charged leptons) in
equilibrium. It includes the one-loop matter potential contribution
($\mathcal{O}(G)$) and the two-loop ($\mathcal{O}(G^2)$)  retarded
self-energy. The ``index of refraction''\cite{notzold}  is
determined by the matter potential and the real part of the
space-time Fourier transform of the retarded self-energy. The
relaxation rates of the quasiparticle excitations are determined by
its imaginary part. The non-equilibrium effective action  leads to
Langevin-like equations of motion for the fields with a noise term
determined by the correlations of the bath, it features a Gaussian
probability distribution but is \emph{colored}. The noise
correlators and the self-energy fulfill a generalized
fluctuation-dissipation relation. }

\item{We obtain expressions for the dispersion relations and mixing
angles in medium which are of the same form as in the case for
neutrinos. The relaxation rates for the two types of quasiparticles
are given by \bea \Gamma_1(k)  & = &   \Gamma_{ee}(k)
\cos^2\theta_m(k)+\Gamma_{\mu\mu}(k)\sin^2\theta_m(k)
\label{Gama1ee} \\\Gamma_2(k)  & = &   \Gamma_{\mu\mu}(k)
\cos^2\theta_m(k)+\Gamma_{ee}(k)\sin^2\theta_m(k)
\label{Gama2mumu}\eea where $\Gamma_{\alpha\alpha}(k)$ are the
relaxation rates of the flavor fields in \emph{absence} of mixing,
and $\theta_m(k)$ is the mixing angle in the medium. }

\item{A  Weisskopf-Wigner description of the long time
dynamics in terms of an effective \emph{non-hermitian Hamiltonian}
is obtained. Although this effective description accurately captures
the asymptotic long time dynamics of the expectation value of the
fields in weak coupling, it \emph{does not} describe the process of
equilibration. }

\item{For long time $>> \Gamma^{-1}_{1,2}$ the two point correlation
function of fields becomes time translational invariant reflecting
the approach to equilibrium. The one-body density matrix reaches its
equilibrium form at long time, in perturbation theory it is
\emph{nearly  diagonal in the basis of   eigenstates  of an
effective   Hamiltonian that includes self-energy corrections in the
medium}, with perturbatively small off-diagonal corrections in this
basis. The diagonal components are determined by the  distribution
function of   eigenstates of this in-medium effective Hamiltonian. }

\item{ These results apply to the case of sterile neutrinos with
 modifications to the dispersion relations and relaxation
rates arising from simple ``sterility'' conditions. ``Sterile
neutrinos'' equilibrate with the bath as a consequence of
active-sterile mixing\cite{dm,aba,foot}. }

\end{itemize}

In section (\ref{sec:model}) we introduce the model and discuss the
ambiguities in defining flavor Fock operators, states and
distribution functions. In section (\ref{sec:noneLeff}) we obtain
the reduced density matrix, the non-equilibrium effective action and
the Langevin-like equations of motion for the expectation value of
the fields. In section (\ref{sec:langevin}) we provide the general
solution of the Langevin equation. In section (\ref{sec:quasi}) we
obtain  the dispersion relations, mixing angles and decay rates of
  quasiparticle modes in the medium. In this section an
effective Weisskopf-Wigner description of the long time dynamics is
derived. In section (\ref{sec:equilibration}) we study the approach
to equilibrium in terms of the  one-body density matrix. In this
section we discuss the consequences   for ``sterile neutrinos''.
Section (\ref{sec:conclu}) summarizes our conclusions.

\section{The model}\label{sec:model} We consider a model of mesons
with two flavors $e\;,\mu$ in interaction with a ``charged current''
  denoted    $W$ and a ``flavor lepton'' $\chi_\alpha$ modeling
the charged current interactions in the electroweak (EW) model. In
terms of field doublets

\be \Phi =  \Bigg( \begin{array}{c}
       \phi_e \\
              \phi_\mu \\
            \end{array} \Bigg) ~~;~~ X =  \Bigg( \begin{array}{c}
       \chi_e \\
              \chi_\mu \\
            \end{array} \Bigg) \label{doublets}\ee the Lagrangian
            density is
\be \mathcal{L}  =  \frac{1}{2} \left\{ \partial_{\mu} \Phi^T
\partial^{\mu} \Phi -   \Phi^T  \mathbb{M}^2    \Phi \right\}+  \mathcal{L}_0[W,\chi]+G \,W\,\Phi^T \cdot X
 +G\phi^2_e \chi^2_e +G\phi^2_{\mu} \chi^2_{\mu} \label{lagra} \ee where the mass matrix
 is given by

\be \mathbb{M}^2  = \left( \begin{array}{cc}
                                   M^2_{ee} & M^2_{e\mu} \\
                                   M^2_{e\mu} & M^2_{\mu\mu} \\
                                 \end{array} \right)
                                 \label{massmatrices} \ee where
 $\mathcal{L}_0[W,\chi]$ is the free field Lagrangian density for $W,\chi$
 which need not be specified. The mesons $\phi_{e,\mu}$ play the
 role of the flavored neutrinos, $\chi_{e,\mu}$ the role of the
 charged leptons and $W$   a  charged  current, for example
 the proton-neutron current
 $\overline{p}\gamma^\mu(1-g_A\gamma_5)n$ or a similar quark
 current. The coupling $G$ plays the role of $G_F$. As it will be seen below, we do not need to specify the
 precise form, only the spectral properties of the correlation
 function of this current are necessary.

Passing from the flavor to the mass basis for the fields
$\phi_{e,\mu}$ by an orthogonal transformation $\Phi = U(\theta)\,
\varphi$

\be   \left(\begin{array}{c}
      \phi_e \\
      \phi_\mu\\
    \end{array}\right) =  U(\theta) ~\Bigg(\begin{array}{c}
                                               \varphi_1 \\
                                               \varphi_2\\
                                             \end{array}\Bigg)~~;~~U(\theta)
                                             = \Bigg( \begin{array}{cc}
                           \cos\theta & \sin\theta \\
                           -\sin\theta & \cos\theta \\
                         \end{array} \Bigg) \label{trafo} \ee where
the orthogonal matrix $U(\theta)$ diagonalizes the mass matrix
$\mathbb{M}^2$, namely

\be U^{-1}(\theta)\,\mathbb{M}^2 \, U(\theta) = \Bigg(
\begin{array}{cc}
                                         M^2_1 &0 \\
                                         0 & M^2_2 \\
                                       \end{array} \Bigg)
                                       \label{diagM} \ee

In the flavor basis   $\mathbb{M} $ can be written   as follows

\be \mathbb{M}^2  = \overline{M}^{\,2}\,\mathds{1}+\frac{\delta
M^2}{2} \left(\begin{array}{cc}
                                                                -\cos 2\theta & \sin2\theta \\
                                                                \sin 2\theta & \cos 2\theta \\
                                                              \end{array}
\right) \label{massmatx2}\ee where we introduced

\be \overline{M}^{\,2} =\frac{1}{2}(M^2_1+M^2_2)~~;~~ \delta M^2 =
M^2_2-M^2_1 \label{MbarDelM}\,. \ee

\subsection{Mass and flavor states:}\label{flavmass}

It is convenient to take the spatial Fourier transform of the fields
$\phi_{\alpha};\varphi_i$ and their canonical momenta $\pi_{\alpha}=
\dot{\phi}_\alpha;\upsilon_i = \dot{\varphi}_i$  with $\alpha=e,\mu$
and $i=1,2$ and write (at t=0),

\bea \phi_{\alpha}(\vec{x}) & = &  \frac{1}{\sqrt{V}}\sum_{\vec{k}}
\phi_{\alpha,\vec{k}}\,e^{i\vec{k}\cdot\vec{x}} ~~;~~
\varphi_{i}(\vec{x})   =    \frac{1}{\sqrt{V}}\sum_{\vec{k}}
\varphi_{i,\vec{k}}\,e^{i\vec{k}\cdot\vec{x}}\nonumber
\\ \pi_{\alpha}(\vec{x}) & = &  \frac{1}{\sqrt{V}}\sum_{\vec{k}}
\pi_{\alpha,\vec{k}}\,e^{i\vec{k}\cdot\vec{x}}~~;~~\upsilon_i(\vec{x})
=\frac{1}{\sqrt{V}}\sum_{\vec{k}}
\upsilon_{i,\vec{k}}\,e^{i\vec{k}\cdot\vec{x}}\label{FT}\eea in
these expressions we have denoted the spatial Fourier transforms
with the same name to avoid cluttering of notation but it is clear
from the argument which variable is used. The free field Fock states
associated with mass eigenstates are obtained by writing the fields
which define the mass basis $\varphi_i$ in terms of creation and
annihilation operators,

\be  \varphi_{i,\vec{k}}    =   \frac{1}{\sqrt{2\omega_i(k)}} \left[
a_{i,\vec{k}}+a^{\dagger}_{i,-\vec{k}} \right]~~;~~
\upsilon_{i,\vec{k}}
  =    \frac{-i\omega_i(k)}{\sqrt{2\omega_i(k)}} \left[
a_{i,\vec{k}}-a^{\dagger}_{i,-\vec{k}} \right] \label{aadaggermass}
\ee  with \be \omega_i(k) = \sqrt{k^2+M^2_i} ~~;~~i=1,2
\label{omegais}\ee The annihilation ($a_{i,\vec{k}}$) and creation
($a^\dagger_{i,\vk}$) operators obey the usual canonical commutation
relations, and the free Hamiltonian in the mass basis is the usual
sum of independent harmonic oscillators with frequencies
$\omega_i(k)$. One can, in principle, \emph{define} annihilation and
creation operators associated with the flavor fields
$a_{\alpha,\vk},a^\dagger_{\alpha,\vk}$ respectively in a similar
manner

\be  \phi_{\alpha,\vec{k}}   =    \frac{1}{\sqrt{2\Omega_\alpha(k)}}
\left[ a_{\alpha,\vec{k}}+a^{\dagger}_{\alpha,-\vec{k}} \right]
~~;~~ \pi_{\alpha,\vec{k}}    =
\frac{-i\Omega_\alpha(k)}{\sqrt{2\Omega_\alpha(k)}} \left[
a_{\alpha,\vec{k}}-a^{\dagger}_{\alpha,-\vec{k}} \right]
\label{aadaggerflav} \ee  with the annihilation
($a_{\alpha,\vec{k}}$) and creation ($a^\dagger_{\alpha,\vec{k}}$)
operators obeying the usual canonical commutation relations.
However, unlike the case for the mass eigenstates, the frequencies
$\Omega_{\alpha}(k)$ are \emph{arbitrary}. \emph{Any choice} of
these frequencies furnishes a \emph{different} Fock representation,
therefore there is an intrinsic ambiguity in defining Fock creation
and annihilation operators for the \emph{flavor} fields since these
do not have a definite mass. In references\cite{blasone,fujii,ji} a
particular assignment of masses has been made, but any other is
equally suitable. The orthogonal transformation between the flavor
and mass fields eqn. (\ref{trafo}), leads to the following relations
between the flavor and mass Fock operators,

\bea  a_{e,\vk}  & = &     \cos\theta
\Bigg[a_{1,\vk}\,\mathbf{A}_{e,1}(k)+a^{\dagger}_{1,-\vk}\,\mathbf{B}_{e,1}(k)
\Bigg]+\sin\theta
\Bigg[a_{2,\vk}\,\mathbf{A}_{e,2}(k)+a^{\dagger}_{2,-\vk}\,\mathbf{B}_{e,2}(k)
\Bigg]    \label{bogotrafoe} \\a_{\mu,\vk}  & = &  \cos\theta
\Bigg[a_{2,\vk}\,\mathbf{A}_{\mu,2}(k)+a^{\dagger}_{2,-\vk}\,\mathbf{B}_{\mu,2}(k)
\Bigg]-\sin\theta
\Bigg[a_{1,\vk}\,\mathbf{A}_{\mu,1}(k)+a^{\dagger}_{1,-\vk}\,\mathbf{B}_{\mu,1}(k)
\Bigg]   \label{bogotrafomu} \eea  where
$\mathbf{A}_{\alpha,i},\mathbf{B}_{\alpha,i}$ are the generalized
Bogoliubov coefficients

\be  \mathbf{A}_{\alpha,i}  =   \frac{1}{2}
\Bigg(\sqrt{\frac{\Omega_\alpha(k)}{\omega_i(k)}}+\sqrt{\frac{\omega_i(k)}{\Omega_\alpha(k)}}
\Bigg) ~~;~~ \mathbf{B}_{\alpha,i}   =  \frac{1}{2}
\Bigg(\sqrt{\frac{\Omega_\alpha(k)}{\omega_i(k)}}-\sqrt{\frac{\omega_i(k)}{\Omega_\alpha(k)}}
\Bigg)\,. \label{bigAB}\ee  These coefficients obey the condition

\be  \left(\mathbf{A}^2_{\alpha,i}-\mathbf{B}^2_{\alpha,i}\right) =
1 \label{unicon}\ee which guarantees that the transformation between
mass and flavor Fock operators is formally unitary   and both sets
of operators obey the canonical commutation relations for \emph{any}
choice of the frequencies $\Omega_{\alpha}(k)$. Neglecting the
interactions, the ground state $|0>$ of the Hamiltonian   is the
vacuum annihilated by the Fock annihilation operators of the mass
basis, \be a_{i,\vk}|0> = 0 ~~ \mathrm{for ~all}~ i=1,2 \,,\vk
\,.\label{vac}\ee

In particular the number of \emph{flavor} Fock quanta in the
non-interacting ground state, which is the \emph{vacuum} of mass
eigenstates is

\bea <0|a^{\dagger}_{e,\vk}\,a_{e,\vk}|0> & = & \cos^2\theta \,
\frac{\Big[\Omega_e(k)-\omega_1(k)\Big]^2}{4\,\Omega_e(k)\,\omega_1(k)}
 +\sin^2\theta \,\frac{\Big[\Omega_e(k)-\omega_2(k)\Big]^2}{4\,\Omega_e(k)\,\omega_2(k)}\label{nume}\\
<0|a^{\dagger}_{\mu,\vk}\,a_{\mu,\vk}|0> & = & \cos^2\theta \,
\frac{\Big[\Omega_\mu(k)-\omega_2(k)\Big]^2}{4\,\Omega_\mu(k)\,\omega_2(k)}+\sin^2\theta
\,
\frac{\Big[\Omega_\mu(k)-\omega_1(k)\Big]^2}{4\,\Omega_\mu(k)\,\omega_1(k)}
\label{numu}\eea namely the non-interacting ground state (the vacuum
of mass eigenstates) is a \emph{condensate} of ``flavor''
states\cite{blasone,fujii,ji} with an average number of ``flavored
particles'' that depends on the arbitrary frequencies
$\Omega_{\alpha}(k)$. Therefore these ``flavor occupation numbers''
or ``flavor distribution functions'' are \emph{not}  suitable
quantities to study equilibration.

\emph{Assuming} that $\Omega_{\alpha}(k) \rightarrow k$ when
$k\rightarrow \infty$, in the high energy limit $\mathbf{A}
\rightarrow 1~;~ \mathbf{B} \rightarrow 0$ and in this high energy
limit

\be  a_{e,\vk}    \approx     \cos\theta \,
 a_{1,\vk} +\sin\theta \,
  a_{2,\vk}   ~~;~~ \\a_{\mu,\vk}    \approx    \cos\theta
  a_{2,\vk} -\sin\theta
 a_{1,\vk}  \label{bogotrafomuhik} \ee

Therefore, under the assumption that the arbitrary frequencies
$\Omega_{\alpha}(k) \rightarrow k$ in the high energy limit, there
is an approximate identification between Fock states in the mass and
flavor basis in this limit. However, such identification is only
\emph{approximate} and only available in the asymptotic regime of
large momentum, but becomes ambiguous for arbitrary momenta. In
summary the definition of flavor Fock states is ambiguous,   the
ambiguity may \emph{only} be approximately resolved in the very high
energy limit, but it is clear that there is no unique definition of
a flavor \emph{distribution function} which is valid for all values
of momentum $k$ and that can serve as a definite yardstick to study
equilibration. Even the non-interacting ground state features an
arbitrary number of flavor Fock quanta depending on the arbitrary
choice of the frequencies $\Omega_{\alpha}(k)$ in the definition of
the flavor Fock operators. This is not a consequence of the meson
model but a \emph{general} feature in the case of mixed fields with
similar ambiguities in the spinor case\cite{chargedlepton}.

We emphasize that while the flavor Fock operators are
\emph{ambiguous} and not uniquely defined, there is no ambiguity in
the flavor \emph{fields} $\phi_\alpha$ which are  related to the
mass fields $\varphi_i$ via the unitary transformation
(\ref{trafo}). While there is no unambiguous definition of the
flavor number operator or distribution function, there is an
unambiguous number operator for the Fock quanta in the mass basis $
N_i(k) = a^\dagger_{i,\vk}\, a_{i,\vk}$, whose expectation value is
the distribution function for mass Fock states.

\section{Reduced density matrix and  non-equilibrium effective
action}\label{sec:noneLeff}

Our goal is to study the equilibration of neutrinos with a bath of
hadrons or quarks and charged leptons in thermal equilibrium at high
temperature. This setting describes the thermalization of neutrinos
in the early Universe prior to BBN, for temperatures $T \gtrsim 3\,
\textrm{MeV}$\cite{dolgovrev,steigman,book3}.

 We focus on the
dynamics of the ``system fields'', either the flavor fields
$\phi_\alpha$ or alternatively   the mass fields $\varphi_i$. The
strategy is to consider the time evolved full density matrix and
trace over the bath degrees of freedom $\chi,W$. It is convenient to
write the Lagrangian density (\ref{lagra}) as
\begin{equation}\label{lagra2}
{\cal L}[\phi_\alpha,\chi_\alpha,W]= {\cal L}_{0}[\phi]+{\cal
L}_{0}[W,\chi]+G\phi_\alpha \mathcal{O}_\alpha+ G \phi^2_\alpha
\chi^2_\alpha
\end{equation} with an implicit sum over the flavor label $\alpha=e,\mu$, where
\be \mathcal{O}_\alpha =   \chi_\alpha\,W \,.\label{calO}\ee
${\mathcal L}_{0}[\cdots]$  are the free  Lagrangian densities for
the   fields $\phi_\alpha,\chi_\alpha,W$ respectively.  The fields
$\phi_\alpha$ are considered as the ``system''  and the fields
$\chi_\alpha,W$ are treated as a bath in thermal equilibrium at a
temperature $T\equiv 1/\beta$. We consider a factorized initial
density matrix at a time $t_i =0$ of the form \begin{equation}
\hat{\rho}(0) = \hat{\rho}_{\phi}(0) \otimes e^{-\beta\,H_0[\chi,W]}
\label{inidensmtx}
\end{equation} where $H_0[\chi,W]$ is   Hamiltonian for
the fields $\chi,W$. Although this factorized form of the initial
density matrix leads to initial transient dynamics, we are
interested in the long time dynamics, in particular in the   long
time limit. The  bath fields $\chi_\alpha,W$  will be ``integrated
out'' yielding a reduced density matrix for the fields $\phi_\alpha$
in terms of an effective real-time functional, known as the
influence functional\cite{feyver} in the theory of quantum brownian
motion. The reduced density matrix can be represented by a path
integral in terms of the non-equilibrium effective action that
includes the influence functional.  This method has been used
extensively to study quantum brownian motion\cite{feyver,leggett},
and quantum kinetics\cite{boyalamo,hoboydavey}.

In the flavor field basis the matrix elements of
$\hat{\rho}_{\phi}(0)$ are given by
\begin{equation}
\langle \phi_\alpha |\hat{\rho}_{\phi}(0) | \phi'_{\beta}\rangle =
\rho_{\phi;0}(\phi_\alpha ;\phi'_\beta)
\end{equation} or alternatively in the mass field basis
\begin{equation}
\langle \varphi_i |\hat{\rho}_{\varphi}(0) | \varphi'_j\rangle =
\rho_{\varphi;0}(\varphi_i ;\varphi'_j)\,.
\end{equation}
The time evolution of the initial density matrix is given by
\begin{equation}\label{rhooft} \hat{\rho}(t_f)=
e^{-iH(t_f-t_i)}\hat{\rho}(t_i)e^{iH(t_f-t_i)}\,,
\end{equation} where the total Hamiltonian $H$ is
\begin{equation}\label{hami}
H=H_{0}[\phi] + H_{0}[\chi,W]+H_I[\phi,\chi,W]\,.
\end{equation} The calculation of correlation functions is facilitated by
introducing currents coupled to the different fields. Furthermore
since each time evolution operator in eqn. (\ref{rhooft}) will be
represented as a path integral, we introduce different sources for
 forward and backward time evolution operators, referred to as
 $J^{+},J^{-}$ respectively.  The
forward and backward time evolution operators in presence of sources
are $U(t_f,t_i;J^+)$, $U^{-1}(t_f,t_i,J^{-})$ respectively.

We will only study  correlation functions of the  ``system'' fields
$\phi$ (or $\varphi$ in the mass basis), therefore we carry out the
trace over the $\chi$ and $W$ degrees of freedom. Since the currents
$J^{\pm}$ allow us to obtain the correlation functions for any
arbitrary time by simple variational derivatives with respect to
these sources, we take $t_f\rightarrow \infty$ without loss of
generality. The non-equilibrium generating functional is  given
by\cite{boyalamo,hoboydavey}

\be \label{noneqgen}{\cal Z}[j^+,j^-] =
\mathrm{Tr}U(\infty,t_i;J^+)\hat{\rho}(t_i)U^{-1}(\infty,t_i,J^{-})\,.
\ee

Where  $J^{\pm}$ stand collectively for all the sources coupled to
different fields. Functional derivatives with respect to the sources
$J^+$ generate the time ordered correlation functions, those with
respect to $J^-$ generate the anti-time ordered correlation
functions and mixed functional derivatives with respect to $J^+,J^-$
generate mixed correlation functions. Each one of the time evolution
operators in the generating functional (\ref{noneqgen}) can be
written in terms of a path integral: the time evolution operator
$U(\infty,t_i;J^+)$ involves a path integral \emph{forward} in time
from $t_i$ to $t=\infty$ in presence of sources $J^+$, while the
inverse time evolution operator $U^{-1}(\infty,t_i,J^{-})$ involves
a path integral \emph{backwards} in time from $t=\infty$ back to
$t_i$ in presence of sources $J^-$. Finally the equilibrium density
matrix for the bath $e^{-\beta\,H_0 [\chi,W]} $ can be written as a
path integral along imaginary time with sources $J^{\beta}$.
Therefore the path integral form of the generating functional
(\ref{noneqgen}) is given by

\be {\cal Z}[j^+,j^-] =
 \int D\Phi_i    D\Phi'_i\,
\rho_{\Phi,i}(\Phi_i;\Phi'_i) \int
 {\cal D}\Phi^{\pm}
 {\cal D} \chi^{\pm}  {\cal D} W^{\pm}{\cal D}
\chi^{\beta}{\cal D} W^{\beta}\,
e^{iS[\Phi^{\pm},\chi^{\pm},W^\pm;J^{\pm}_{\Phi};J^{\pm}_{\chi};J^\pm_W]}
 \label{pathint}\ee

 \noindent with the boundary conditions $\Phi^+(\vec
 {x},t_i)=\Phi_i(\vec{x})\,;\,\Phi^-(\vec
 {x},t_i)=\Phi^{'}_i(\vec{x})$. The trace over the bath fields
 $\chi,W$ is performed with the usual periodic boundary conditions
 in Euclidean time.

\begin{figure}[ht!]
\begin{center}
\includegraphics[height=2in,width=4in,keepaspectratio=true]{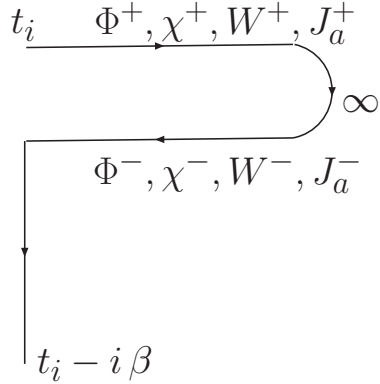}
\caption{Contour in time for the non-equilibrium path integral
representation.  } \label{fig:ctp}
\end{center}
\end{figure}

The non-equilibrium  action is given by
 \bea
S[\Phi^{\pm},\chi^{\pm};J^{\pm}_{\Phi};J^{\pm}_{\chi};J^{\pm}_W] & =
& \int_{t_i}^{\infty}dt
d^3x\,\left[\mathcal{L}_{0}(\phi^+)+J^+_{\phi}\phi^+
-\mathcal{L}_{0}(\phi^-)-J^-_{\phi}\phi^-  \right] +\nonumber
\\ &&
\int_{\mathcal{C}}d^4x\Big\{\mathcal{L}_{0}[\chi,W]+
J_{\chi}\chi+J_W\,W +G\,\phi_\alpha\,\mathcal{O}_\alpha
+G\phi^2_\alpha \chi^2_\alpha \Big\} \label{noneqlagradens} \eea

\noindent where $\mathcal{C}$ describes the following contour in the
complex time plane: along the forward branch $(t_i,+\infty)$
 the fields and sources are $\Phi^{+},\chi^+,J^+_{\chi}$,
along the backward branch  $(\infty, t_i)$  the fields and sources
are $\Phi^{-},\chi^-,J^-_{\chi}$ and along the Euclidean branch
$(t_i, t_i-i\beta)$  the fields and sources are $\Phi=0;
\chi^{\beta},J^{\beta}_{\chi}$. Along the Euclidean branch the
interaction term vanishes since the initial density matrix for the
field $\chi$ is assumed to be that of thermal equilibrium. This
contour is depicted in fig. (\ref{fig:ctp})

The trace over the degrees of freedom of the $\chi$ field with the
initial equilibrium  density matrix,  entail periodic   boundary
conditions for $\chi,W$ along the contour $\mathcal{C}$. However,
the boundary conditions on the path integrals for the field $\Phi$
are given by

\be \Phi^+(\vec x,t=\infty) =\Phi^-(\vec x,t=\infty)\label{endbc}\ee

\noindent and
\begin{equation}
\Phi^+(\vec x,t=t_i)= \Phi_i(\vec x)  \; \; \; ; \; \; \;
\Phi^-(\vec x,t=t_i)=\Phi'_i(\vec x) \label{condsfi}
\end{equation}

The reason for the different path integrations is that whereas the
$\chi$ and $W$ fields are traced over with an initial thermal
density matrix, the initial density matrix for the $\Phi$ field will
be specified later as part of the initial value problem. The path
integral over $\chi,W$ leads to the influence functional for
$\Phi^{\pm}$\cite{feyver}.

Because we are not interested in the correlation functions of the
bath fields but only those of the ``system'' fields, we set the
external c-number currents $J_\chi=0;J_W=0$. Insofar as the bath
fields are concerned, the system fields $\Phi$ act as an external
c-number source, and tracing over the bath fields leads to

\be\int {\cal D} \chi^{\pm} {\cal D} W^{\pm} {\cal D}\chi^{\beta}
{\cal D} W^\beta\,
e^{i\int_{\mathcal{C}}d^4x\Big\{\mathcal{L}_{0}[\chi,W]+
G\,\phi_\alpha \mathcal{O}_\alpha  +G\phi^2_\alpha \chi^2_\alpha
 \Big\}} = \Big\langle e^{iG \int_{\mathcal{C}}d^4x \phi_\alpha
\mathcal{O}_\alpha+\phi^2_\alpha \chi^2_\alpha}
\Big\rangle_{0}\,\mathrm{Tr}e^{-\beta H^0[\chi,W]}. \label{aver}\ee

The expectation value in the right hand side of eqn. (\ref{aver}) is
in the equilibrium free field density matrix of the fields $\chi,W$.
The path integral can be carried out in perturbation theory and the
result exponentiated to yield the effective action as follows

\bea \Big\langle e^{iG \int_{\mathcal{C}}d^4x \phi_\alpha
\mathcal{O}_\alpha+\phi^2_\alpha \chi^2_\alpha} \Big\rangle_{0} = &&
1+    iG
 \int_{\mathcal{C}}d^4x \Big\{
 \phi_\alpha(x)\,\Big\langle\mathcal{O}_\alpha(x)\Big\rangle_{0}+ \phi^2_\alpha(x) \,\Big\langle\chi^2_\alpha(x)\Big\rangle_{0}\Big\}+ \nonumber\\
  && \frac{(iG)^2}{2}\int_{\mathcal{C}}d^4x
 \int_{\mathcal{C}}d^4x'\phi_\alpha(x)\phi_\beta(x')\Big\langle
 \mathcal{O}_\alpha(x)\mathcal{O}_\beta(x')\Big\rangle_{0}+\mathcal{O}(G^3)
 \eea

This is the usual expansion of the exponential of the connected
correlation functions, therefore this series is identified with

\be \Big\langle e^{iG
\int_{\mathcal{C}}d^4x \phi_\alpha \mathcal{O}_\alpha+\phi^2_\alpha \chi^2_\alpha} \Big\rangle_{0} = e^{i\,L_{if}[\phi^+,\phi^-]
}\;,
 \ee where $L_{if}[\phi^+,\phi^-]$ is the
\emph{influence functional}\cite{feyver}, and $\langle \cdots
\rangle_0$ stand for expectation values in the bath in equilibrium.
For $\Big\langle\chi_\alpha(x)W(x)\Big\rangle_0 =0$ the influence
functional is given by

\be L_{if}[\phi^+,\phi^-] = G \int_{\mathcal{C}}d^4x
\phi^2_\alpha(x) \,\Big\langle\chi^2_\alpha(x)\Big\rangle_{0}+
 i\frac{G^2}{2}\int_{\mathcal{C}}d^4x
 \int_{\mathcal{C}}d^4x'\phi_\alpha(x)\phi_\beta(x')\langle
 \mathcal{O}_\alpha(x)\mathcal{O}_\beta(x')\rangle_{0}+\mathcal{O}(G^3)\,. \label{Lif}\ee

In the above result we have neglected second order contributions of
the form $G^2 \phi^4_\alpha$. These non-linear contributions give
rise to interactions between the quasiparticles and \emph{will be
neglected} in this article. Here we are primarily concerned with
establishing the general properties of the quasiparticles and their
equilibration with the bath and not with their mutual interaction.
As in the case of mixed neutrinos, the inclusion of a ``neutrino''
background may lead to the phenomenon of non-linear
synchronization\cite{pantaleone,samuel,synchro2}, but the study of
this phenomenon is beyond the realm of this article.

We focus solely on the non-equilibrium effective action up to
quadratic order in the ``neutrino fields'', from which we  extract
the   dispersion relations, relaxation rates and the  approach to
equilibrium with the bath of the quasiparticle modes in the medium.

The integrals along the contour $\mathcal{C}$ stand for the
following expressions:

\be G\int_{\mathcal{C}}d^4x \phi^2_\alpha(x)
\,\Big\langle\chi^2_\alpha(x)\Big\rangle_{0} = V_{\alpha\alpha} ~
\int
 d^3x \int_{t_i}^{\infty}dt  \left[\phi^{2\,+}_\alpha(x)-\phi^{2\,-}_\alpha(x) \right]
 \ee where $V_{\alpha\alpha}$ are the ``matter potentials'' which are
 independent of position under the assumption of translational
 invariance, and time independent under the assumption that the bath
 is in equilibrium, and

\bea \int_{\mathcal{C}}d^4x
 \int_{\mathcal{C}}d^4x'\phi_\alpha(x)\phi_\beta(x')\langle
 \mathcal{O}_\alpha(x)\mathcal{O}_\beta(x')\rangle_{0} = &&\int
 d^3x \int_{t_i}^{\infty}dt \int
 d^3x' \int_{t_i}^{\infty}dt' \Big[ \phi^+_\alpha(x)\phi^+_\beta(x')\langle
 \mathcal{O}^+_\alpha(x)\mathcal{O}^+_\beta(x')\rangle_{0}
 \nonumber\\+&&
 \phi^-_\alpha(x)\phi^-_\beta(x')\langle
 \mathcal{O}^-_\alpha(x)\mathcal{O}^-_\beta(x')\rangle_{0}
 -  \phi^+_\alpha(x)\phi^-_\beta(x')\langle
 \mathcal{O}^+_\alpha(x)\mathcal{O}^-_\beta(x')\rangle_{0}\nonumber \\ -&&
 \phi^-_\alpha(x)\phi^+_\beta(x')\langle
 \mathcal{O}^-_\alpha(x)\mathcal{O}^+_\beta(x')\rangle_{0}\Big]\eea

Since the expectation values above are computed in a thermal
equilibrium translational invariant density matrix, it is convenient
to  introduce the spatial Fourier transform of the composite
operator ${\cal O}$ in a spatial volume $V$ as
\begin{equation}
{\cal O}_{\alpha,\vec k}(t) = \frac{1}{\sqrt{V}} \int d^3x e^{i \vec
k \cdot \vec x}
 {\cal O}_\alpha(\vec x,t)
\label{spatialFT}
\end{equation}
 in terms of which we obtain
following the correlation functions
\begin{eqnarray}
 && \langle \mathcal {O}^-_{\alpha,\vec k}(t) {\cal O}^+_{\beta,-\vec k}(t')\rangle =
\mathrm{Tr}\,{\cal O}_{\beta,-\vec k}(t')\,
e^{-\beta\,H^0[\chi,W]}\, \mathcal{O}_{\alpha,\vec k}(t)=
\mathcal{G}^>_{\alpha \beta}(k;t-t') \equiv {\cal G}^{-+}_{\alpha
\beta}(k;t-t')\label{ggreat}
\\&&
 \langle {\cal O}^+_{\alpha,\vec k}(t) {\cal O}^{-}_{\beta,-\vec k}(t')\rangle=
\mathrm{Tr}\,\mathcal{O}_{\alpha,\vec k}(t) \,
e^{-\beta\,H_{\chi}}\, {\cal O}_{\beta,-\vec k}(t') ={\cal
G}^<_{\alpha \beta}(k;t-t') \equiv {\cal G}^{+-}_{\alpha
\beta}(k;t-t')= {\cal G}^{-+}_{\beta,\alpha}(k;t'-t)\label{lesser}
\\&& \langle {\cal O}^+_{\alpha,\vec k}(t) {\cal O}^+_{\beta,-\vec
k}(t')\rangle = {\cal G}^>_{\alpha \beta}(k;t-t')\Theta(t-t')+ {\cal
G}^<_{\alpha \beta}(k;t-t')\Theta(t'-t)\equiv {\cal G}^{++}_{\alpha
\beta}(k;t-t') \label{timeordered} \\&& \langle {\cal
O}^-_{\alpha,\vec k}(t) {\cal O}^-_{\beta,-\vec k}(t')\rangle =
{\cal G}^>_{\alpha \beta}(k;t-t')\Theta(t'-t)+ {\cal G}^<_{\alpha
\beta}(k;t-t')\Theta(t-t') = {\cal G}^{--}_{\alpha
\beta}(k;t-t')\label{antitimeordered}
\end{eqnarray}

The time evolution of the operators is determined by the Heisenberg
picture of $H_{0}[\chi,W]$. Because the density matrix for the bath
is in equilibrium, the correlation functions above are solely
functions of the time difference as made explicit in the expressions
above. These correlation functions are not independent, but obey

\begin{equation}\label{const}
\mathcal{ G}^{++}_{\alpha \beta}(k;t,t') + \mathcal{ G}^{--}_{\alpha
\beta}(k;t,t')-\mathcal{G}^{-+}_{\alpha
\beta}(k;t,t')-\mathcal{G}^{+-}_{\alpha \beta}(k;t,t')=0
\end{equation}

The correlation function $\mathcal{G}^{>}_{\alpha \beta}$ up to
lowest order in the coupling $G$ is given by

\be  \mathcal{G}^>_{\alpha \beta}(k;t-t')   =   \int
\frac{d^3p}{(2\pi)^3} \langle W_{\vec{p}+\vk}(t)
W_{-\vec{p}-\vk}(t')\rangle \langle
\chi_{\vec{p},\alpha}(t)\chi_{-\vec{p},\beta}(t')\rangle
\label{corre} \ee where the expectation value is in the free field
equilibrium density matrix of the respective fields. This
correlation function is diagonal in the flavor basis and this
entails that all the Green's functions
(\ref{ggreat}-\ref{antitimeordered}) are diagonal in the flavor
basis.

The non-equilibrium effective action yields the time evolution of
the reduced density matrix, it is given by

\be\label{noneqeff} L_{eff}[\phi^+,\phi^-] = \int_{t_i}^{\infty}dt
d^3x\,\left[\mathcal{L}_{0}(\phi^+) -\mathcal{L}_{0}(\phi^-)
\right] + L_{if}[\phi^+,\phi^-]\ee

\noindent where we have set the sources $J^\pm$ for the fields
$\phi^\pm$ to zero.

In what follows we take $t_i =0$ without loss of generality since
(i) for $t > t_i$ the total Hamiltonian is time independent and the
correlations will be solely functions of $t-t_i$, and (ii) we will
be ultimately interested in the limit $t \gg t_i$ when all transient
phenomena has relaxed.  Adapting the methods presented  in ref.
\cite{hoboydavey}  to account for the  matrix structure of the
effective action,   introducing  the spatial Fourier transform of
the fields $\phi^\pm$ defined as in eqn. (\ref{spatialFT}) and   the
matrix of the matter potentials \be \mathds{V} = \left(%
\begin{array}{cc}
  V_{ee} & 0 \\
  0 & V_{\mu\mu} \\
\end{array}%
\right) \label{matV}\ee      we find
\begin{eqnarray}\label{influfunc} iL_{eff}[\phi^+,\phi^-] & = &
\sum_{\vec k}\Bigg\{ \frac{i}{2} \int_0^{\infty} dt
\Big[\dot{\phi}^+_{\alpha,\vec k}(t)\dot{\phi}^+_{\alpha,-\vec
k}(t)- \phi^+_{\alpha,\vec k}(t)(k^2\,\delta_{\alpha \beta}+
\mathbb{M}^2_{\alpha
\beta}+\mathds{V}_{\alpha\beta}) \phi^+_{\beta,-\vec k}(t)  \nonumber \\
& &
 -\dot{\phi}^-_{\alpha,\vec k}(t)\dot{\phi}^-_{\alpha,-\vec k}(t)+\phi^-_{\alpha,\vec k}(t)(k^2\,\delta_{\alpha \beta}+ \mathbb{M}^2_{\alpha \beta}+\mathds{V}_{\alpha\beta})
  \phi^-_{\beta,-\vec k}(t) \Big]   \nonumber \\
&   &  - \frac{G^2}{2} \int_0^{\infty} dt \int_0^{\infty} dt' \left[
\phi^+_{\alpha,\vec k}(t) {\cal G}^{++}_{\alpha
\beta}(k;t,t')\phi^+_{\beta,-\vec k}(t')+ \phi^-_{\alpha,\vec
k}(t){\cal G}^{--}_{\alpha \beta}(k;t,t') \phi^-_{\beta,-\vec k}(t') \right. \nonumber \\
&& \left. -\phi^+_{\alpha,\vec k}(t){\cal G}^{+-}_{\alpha
\beta}(k;t,t')\phi^-_{\beta,-\vec k}(t')- \phi^-_{\alpha,\vec
k}(t){\cal G}^{-+}_{\alpha \beta}(k;t,t')\phi^+_{\beta,-\vec
k}(t')\right] \Bigg\}
\end{eqnarray}

The ``matter potentials'' $V_{\alpha\alpha}$ play the role of the
index of refraction correction to the dispersion
relations\cite{notzold} and is of first order in the coupling $G$
whereas the contributions that involve $\mathcal{G}$ are of order
$G^2$.   As it will become clear below, it is more convenient to
introduce the Wigner center of mass and relative variables \be
\Psi_\alpha(\vec x,t)   =   \frac{1}{2} \left(\phi^+_\alpha(\vec
x,t) + \phi^-_\alpha(\vec x,t) \right) \; \; ; \; \; R_\alpha(\vec
x,t) = \left(\phi^+_\alpha(\vec x,t) - \phi^-_\alpha(\vec x,t)
\right) \label{wigvars} \ee and the Wigner transform of the initial
density matrix for the $\phi$ fields

\begin{equation}
{\cal W}(\Psi^0 ; \Pi^0) = \int D R_{0,\alpha} e^{-i\int d^3x
\Pi_{0,\alpha}(\vec x) R_{0,\alpha}(\vec x)}
\rho(\Psi^0+\frac{R^0}{2};\Psi^0-\frac{R^0}{2}) \label{Wig}\ee with
the inverse transform

\be \rho(\Psi^0+\frac{R^0}{2};\Psi^0-\frac{R^0}{2})= \int D
\Pi^0_{\alpha} e^{i\int d^3x \Pi^0_{\alpha}(\vec x)
R^0_{\alpha}(\vec x)}{\cal W}(\Psi^0 ; \Pi^0)\label{wignerrho}
\end{equation}

The boundary conditions on the $\phi$ path integral given by
(\ref{condsfi}) translate into the following boundary conditions on
the center of mass and relative variables
\begin{equation}
\Psi_\alpha(\vec x,t=0)= \Psi^0_{\alpha} \; \; ; \; \; R_\alpha(\vec
x,t=0)= R^0_{\alpha} \label{bcwig}
\end{equation}

\noindent furthermore, the boundary condition (\ref{endbc}) yields
the following boundary condition for the relative field \be
\label{Rinfty} R_\alpha(\vec{x},t=\infty)= 0. \ee This observation
will be important in the steps that follow.

The same description applies to the fields in the mass basis. We
will treat both cases on equal footing with the notational
difference that Greek labels $\alpha,\beta$ refer to the flavor and
Latin indices $i,j$ refer to the mass basis.

 In
terms of the spatial Fourier transforms of the center of mass and
relative  variables (\ref{wigvars}) introduced above, integrating by
parts and accounting for the boundary conditions (\ref{bcwig}), the
non-equilibrium effective action (\ref{influfunc}) becomes:
\begin{eqnarray}
iL_{eff}[\Psi,R] & = & \int_0^{\infty} dt \sum_{\vec k} \left\{-i
R_{\alpha,-\vec k}\left( \ddot{\Psi}_{\alpha, \vec k}(t)+
(k^2\delta_{\alpha \beta}+\mathbb{M}^2_{\alpha \beta}+\mathds{V}_{\alpha\beta})\Psi_{\beta,\vk}(t) \right)\right\} \nonumber \\
                 & - & \int_0^{\infty} dt \int_0^{\infty} dt' \sum_{\vec k}\left\{\frac{1}{2}R_{\alpha,-\vec k}(t){\cal K}_{\alpha \beta}(k;t-t')
                  R_{\beta,\vec k}(t')+ R_{\alpha,-\vec k}(t)
\,i\Sigma^R_{\alpha \beta}(k;t-t') \Psi_{\beta,\vec k}(t') \right \} \nonumber \\
& + & i \int d^3x R^0_{\alpha}(\vec x) \dot{\Psi}_\alpha(\vec x,t=0)
\label{efflanwig}
\end{eqnarray}
where the last term arises after the integration by parts in time,
using the boundary conditions (\ref{bcwig}) and (\ref{Rinfty}). The
kernels in the above effective Lagrangian are given by (see eqns.
(\ref{ggreat}-\ref{antitimeordered}))
\begin{eqnarray}
\mathcal{K}_{\alpha \beta}(k;t-t') & = &
\frac{G^2}{2} \left[{\cal G}^>_{\alpha \beta}(k;t-t')+{\cal G}^<_{\alpha \beta}(k;t-t') \right] \label{kernelkappa} \\
i\Sigma^{R}_{\alpha \beta}(k;t-t') & = &  G^2 \left[{\cal
G}^>_{\alpha \beta}(k;t-t')-{\cal G}^<_{\alpha \beta}(k;t-t')
\right]\Theta(t-t') \equiv i\Sigma_{\alpha
\beta}(k;t-t')\Theta(t-t') \label{kernelsigma}
\end{eqnarray}

The term quadratic in the relative variable $R$ can be written in
terms of a stochastic noise as
\begin{eqnarray}
\exp\Big\{-\frac{1}{2} \int dt \int dt' R_{\alpha,-\vec k}(t){\cal
K}_{\alpha \beta}(k;t-t')R_{\beta\vec k}(t')\Big\} = \int {\cal
D}\xi && \exp\Big\{-\frac{1}{2} \int dt \int dt' \, \xi_{\alpha,\vec
k}(t)
{\cal K}^{-1}_{\alpha \beta}(k;t-t')\xi_{\beta,-\vec k}(t')  \nonumber \\
 && +i \int dt \, \xi_{\alpha,-\vec k}(t) R_{\alpha,\vec k}(t)\Big\}
\label{noisefunc}
\end{eqnarray}

The non-equilibrium generating functional can now be written in the
following form
\begin{eqnarray}
{\cal Z}  & = &   \int D \Psi^0 \int D \Pi^0 \int {\cal D} \Psi
{\cal D}R {\cal D}\xi ~~ {\cal W}(\Psi^0;\Pi^0) DR^0
 e^{i \int d^3x R_{0,\alpha}(\vec x) \left(\Pi^0_{\alpha} (\vec x)-\dot{\Psi}_\alpha(\vec x,t=0)\right)}
 {\cal P}[\xi] \label{genefunc} \\
&  & \exp\left\{-i \int_0^{\infty} dt \, R_{\alpha,-\vec k}(t)
\left[ \ddot{\Psi}_{\alpha,\vec k}(t)+(k^2\,\delta_{\alpha
\beta}+\mathbb{M}^2_{\alpha
\beta}+\mathds{V}_{\alpha\beta})\Psi_{\beta,\vec k}(t)+\int dt' \,
\Sigma^{R}_{\alpha \beta}(k;t-t')\Psi_{\beta,\vec k}(t')-\xi_{\alpha,\vec k}(t) \right] \right\} \nonumber \\
{\cal P}[\xi] & = & \exp\left\{-\frac{1}{2} \int_0^{\infty} dt
\int_0^{\infty} dt' ~~ \xi_{\alpha,\vec k}(t) {\cal K}^{-1}_{\alpha
\beta}(k;t-t') \xi_{\beta,-\vec k}(t') \right\} \label{probaxi}
\end{eqnarray} The functional integral over $R^0$ can now be done, resulting in a
functional delta function, that fixes the boundary condition
$\dot{\Psi}_\alpha(\vec x,t=0) = \Pi^0_{\alpha}(\vec x)$. Finally
the path integral over the relative variable can be performed,
leading to a functional delta function and the final form of the
generating functional  given by \begin{eqnarray} {\cal Z}   & = &
\int D \Psi^0   D \Pi^0 \,{\cal W}(\Psi^0;\Pi^0)
 {\cal D} \Psi {\cal D}\xi \, {\cal P}[\xi]\, \times \nonumber \\
  & & \delta\left[
\ddot{\Psi}_{\alpha,\vec k}(t)+(k^2\,\delta_{\alpha
\beta}+\mathbb{M}^2_{\alpha
\beta}+\mathds{V}_{\alpha\beta})\Psi_{\beta,\vec k}(t)+\int_0^{t}
dt' ~\Sigma_{\alpha \beta}(k;t-t')\Psi_{\beta,\vec
k}(t')-\xi_{\alpha,\vec k}(t) \right] \label{deltaprob}
\end{eqnarray}
with the boundary conditions on the path integral on $\Psi$ given by
\begin{equation}
\Psi_\alpha(\vec x,t=0) = \Psi^0_{\alpha}(\vec x) \; \; ; \; \;
\dot{\Psi}_\alpha(\vec x,t=0)= \Pi^0_{\alpha}(\vec x)\,,
\label{bcfin}
\end{equation}  where we have used the definition of $\Sigma^{R}_{\alpha
\beta}(k;t-t')$ in terms of $\Sigma_{\alpha \beta}(k;t-t')$ given in
equation (\ref{kernelsigma}).

The meaning of the above generating functional is the following:
to obtain correlation functions of the center of mass Wigner
variable $\Psi$ we must first find the solution of the classical
{\em stochastic} Langevin equation of motion
\begin{eqnarray}
&& \ddot{\Psi}_{\alpha,\vec k}(t)+(k^2\,\delta_{\alpha
\beta}+\mathbb{M}^2_{\alpha
\beta}+\mathds{V}_{\alpha\beta})\Psi_{\beta,\vec k}(t)+\int_0^t dt'
~ \Sigma_{\alpha \beta}(k;t-t')
\Psi_{\beta,\vec k}(t')=\xi_{\alpha,\vec k}(t) \nonumber \\
&& \Psi_{\alpha,\vec k}(t=0)= \Psi^0_{\alpha,\vec k}~~ ; ~~
\dot{\Psi}_{\alpha,\vec k}(t=0)= \Pi^0_{\alpha,\vec k}
\label{langevin}
\end{eqnarray}
for arbitrary noise term $\xi$ and then average the products of
$\Psi$ over the stochastic noise with the Gaussian probability
distribution ${\cal P}[\xi]$ given by (\ref{probaxi}), and finally
average over the initial configurations $\Psi^0(\vec x); \Pi^0(\vec
x)$ weighted by the Wigner function ${\cal W}(\Psi^0,\Pi^0)$, which
plays the role of an initial
 phase space distribution function.

 Calling
the solution of (\ref{langevin}) $\Psi_{\alpha,\vec
k}(t;\xi;\Psi_i;\Pi_i)$,  the two point correlation function, for
example, is given by
\begin{equation}
\langle \Psi_{\alpha, \vec k}(t) \Psi_{\beta,-\vec k}(t') \rangle =
\frac{ \int {\cal D}[\xi] {\cal P}[\xi] \int D \Psi^0 \int D\Pi^0
~~{\cal W}(\Psi^0;\Pi^0) \Psi_{\alpha,\vec k}(t;\xi;\Psi^0;\Pi^0)
\Psi_{\beta,-\vec k}(t';\xi;\Psi^0;\Pi^0)}{\int {\cal D}[\xi] {\cal
P}[\xi] \int D \Psi^0 \int D\Pi^0 ~~{\cal W}(\Psi^0;\Pi^0)}\,.
\label{expecvalcm}
\end{equation} In computing the averages and using the functional
delta function to constrain the configurations of $\Psi$ to the
solutions of the Langevin equation, there is the Jacobian of the
operator $(d^2/dt^2 + k^2)\delta_{\alpha
\beta}+\mathbb{M}^2+\mathds{V} +\int dt' \Sigma^{ret}(k;t-t')$ which
however, is independent of the field and the noise and cancels
between numerator and denominator in the averages.  There are two
different averages:

\begin{itemize}
\item{ The average over the  stochastic noise term, which up to
this order is Gaussian. We denote the average of a functional
$\mathcal{F}[\xi]$  over the noise with the probability distribution
function $P[\xi]$ given by eqn. (\ref{probaxi}) as

\begin{equation}\label{stocha}
\langle \langle \mathcal{F}  \rangle \rangle \equiv \frac{\int
\mathcal{D}\xi P[\xi] \mathcal{F}[\xi]}{\int \mathcal{D}\xi P[\xi]}.
\end{equation}

Since the noise probability distribution function is Gaussian the
only necessary correlation functions for the noise are given by

\begin{equation}
\langle \langle \xi_{\alpha,\vec{k}}(t)\rangle \rangle =0 \; , \;
\langle \langle
\xi_{\alpha,\vec{k}}(t)\xi_{\beta,\vec{k}'}(t')\rangle \rangle =
{\mathcal K}_{\alpha \beta}(k;t-t')\,\delta^{3}(\vec{k}+\vec{k}')
\label{noisecorrel}
\end{equation}
\noindent and  the higher order correlation functions are obtained
from Wick's theorem as befits a Gaussian distribution function.
Because the noise kernel $\mathcal{K}_{\alpha \beta}(k;t-t')\neq
\delta(t-t')$ the noise is \emph{colored}. }

\item{The average over the initial conditions with the Wigner
distribution function ${\cal W}(\Psi^0,\Pi^0)$ which we denote as

\begin{equation}
\overline{\mathcal{A}[\Psi^0,\Pi^0]} \equiv  \frac{\int D \Psi^0
\int D\Pi^0 ~~{\cal W}(\Psi^0;\Pi^0) \mathcal{A}[\Psi^0,\Pi^0]}{\int
D \Psi^0 \int D\Pi^0 ~~{\cal W}(\Psi^0;\Pi^0) } \label{iniaverage}
\end{equation} }

\end{itemize}

Therefore, the average in the time evolved reduced density matrix
implies \emph{two} distinct averages: an average over the initial
conditions of the system fields and and average over the noise
distribution function. The \emph{total} average is   defined by

\begin{equation}
\langle \cdots \rangle \equiv \overline{\langle \langle
\cdots\rangle \rangle}~~. \label{totave}
\end{equation}

Equal time expectation values and correlation functions are simply
expressed in terms of the  center of mass Wigner variable $\Psi$ as
can be seen as follows: the expectation values of the field

\be \langle \phi^+(\vx,t) \rangle = \mathrm{Tr } \phi(\vx,t)
\rho(t)~~;~~\langle \phi^-(\vx,t) \rangle = \mathrm{Tr } \rho(t)
\phi(\vx,t)\ee  hence the total average (\ref{totave}) is given by
\be \langle \phi(\vx,t) \rangle = \langle \Psi(\vx,t) \rangle\,.
\label{ave}\ee  Similarly the \emph{equal time} correlation
functions obey \be \langle \phi^+(\vx,t)\phi^+(\vx',t) \rangle =
\langle \phi^+(\vx,t)\phi^-(\vx',t) \rangle = \langle
\phi^-(\vx,t)\phi^+(\vx',t) \rangle = \langle
\phi^-(\vx,t)\phi^-(\vx',t) \rangle = \langle
\Psi(\vx,t)\Psi(\vx',t)\rangle \label{psipsi}\,.\ee Therefore the
center of mass variables $\Psi$ contain all the information
necessary to obtain expectation values and equal time correlation
functions.

\subsection{ One body density matrix and equilibration:}\label{equil} We study equilibration by focusing on
the one-body density matrix \be\label{onebody}
\rho_{\alpha\beta}(k;t) = \mathrm{Tr} \rho(0) \phi_{\alpha}(\vk,t)
\phi_{\beta}(-\vk,t) = \mathrm{Tr} \rho(t)\phi_{\alpha}(\vk,0)
\phi_{\beta}(-\vk,0)\ee where \be \rho(t) = e^{-iHt}\rho(0)e^{ iHt}
\label{rhot} \ee is the time evolved density matrix. The time
evolution of the one-body density matrix obeys the Liouville-type
equation \be \label{eqofmotonebody}\frac{d}{dt}
\rho_{\alpha\beta}(t) = -i
\mathrm{Tr}\big[H,\rho(t)\big]\phi_{\alpha}(\vk,0)
\phi_{\beta}(-\vk,0)\,.\ee If the system reaches equilibrium with
the bath at long times, then it is expected that

\be
\big[H,\rho(t)\big]\,\stackrel{t\rightarrow\infty}{\longrightarrow
~~ 0}  \ee Therefore the asymptotically long time limit of the
one-body density matrix yields information on whether the density
matrix is diagonal in the flavor or any other basis. Hence we seek
to obtain \be   \rho_{\alpha\beta}(k;\infty) =
\mathrm{Tr}\rho(\infty)\phi_{\alpha}(\vk,0) \phi_{\beta}(-\vk,0)=
   \langle \Psi_{\alpha \vk}(\infty)
\Psi_{\beta,-\vk}(\infty) \rangle   \,. \label{equione} \ee and to
establish the basis in which it is nearly diagonal. The second
equality in eqn. (\ref{equione}) follows  from eq. (\ref{psipsi}),
and the average is defined by eq.(\ref{totave}). To establish a
guide post, consider   the one-body density matrix for the
\emph{free} ``neutrino fields'' in thermal equilibrium,  for which
the equilibrium density matrix is \be\rho_{eq}= e^{-\beta
H_0[\varphi]} \label{eq}\ee where $H_0[\varphi]$ is the free
``neutrino'' Hamiltonian. This density matrix is \emph{diagonal} in
the basis of mass eigenstates and so is the one-body density matrix
which in the mass basis is given by \be \rho_{ij}(k) = \left(
                     \begin{array}{cc}
                       \frac{1}{2\omega_1(k)}\coth\left[\frac{\beta\omega_1(k)}{2}\right] & 0 \\
                      0  &  \frac{1}{2\omega_2(k)}\coth\left[\frac{\beta\omega_2(k)}{2}\right] \\
                     \end{array}
                   \right) ~~;~~i,j=1,2 \label{rho1mass}\ee therefore in the flavor
                   basis the one-body density matrix is given by
\be \rho_{\alpha\beta}(k) = U(\theta) ~\left(
                     \begin{array}{cc}
                       \frac{1}{2\omega_1(k)}\coth\left[\frac{\beta\omega_1(k)}{2}\right] & 0 \\
                      0  &  \frac{1}{2\omega_2(k)}\coth\left[\frac{\beta\omega_2(k)}{2}\right] \\
                     \end{array}
                   \right) ~U^{-1}(\theta) ~~;~~\alpha,\beta=e,\mu
                   \label{rho1flav}                    \ee This
                   simple example   provides a guide to interpret
the approach to equilibrium. Including interactions there is   a
competition between the mass and flavor basis. The interaction is
diagonal in the flavor basis, while the unperturbed Hamiltonian is
diagonal in the mass basis, this of course is the main physical
reason behind neutrino oscillations. In the presence of
interactions, the correct form of the equilibrium one-body density
matrix can only be obtained from the asymptotic long time limit of
the time-evolved density matrix.

\subsection{Generalized fluctuation-dissipation relation:}

 From the expressions (\ref{kernelkappa}) and (\ref{kernelsigma})   in terms of the Wightmann
functions (\ref{ggreat},\ref{lesser}) which are   averages in the
equilibrium density matrix of the bath fields ($\chi,W$), we
  obtain a dispersive representation for the kernels
$\mathcal{K}_{\alpha \beta}(k;t-t');\Sigma^{R}_{\alpha
\beta}(k;t-t')$.
 This is achieved by   writing the expectation value in
terms of energy eigenstates of the bath,  introducing the identity
in this basis, and using the time evolution of the Heisenberg field
operators to obtain \be G^2~{\cal G}^>_{\alpha \beta}(k;t-t')   =
 \int^{\infty}_{-\infty} d\omega ~ \sigma^>_{\alpha
 \beta}(\vk,\omega)~e^{i\omega(t-t')}
~~;~~ G^2~{\cal G}^<_{\alpha \beta}(k;t-t')   =
\int^{\infty}_{-\infty} d\omega ~ \sigma^<_{\alpha \beta}(\vec
k,\omega)~e^{i\omega(t-t')} \label{specrep} \ee  with the spectral
functions

\begin{eqnarray}
\sigma^>_{\alpha \beta}(\vk,\omega) & = &  \frac{G^2}{\mathcal{Z}_b}
\sum_{m,n}e^{-\beta E_n}
\langle n| {\cal O}_{\alpha,\vec k}(0) |m \rangle \langle m| {\cal O}_{\beta,-\vec k}(0) |n \rangle \, \delta(\omega-(E_n-E_m)) \label{siggreat} \\
\sigma^<_{\alpha \beta}(\vk,\omega) & = &  \frac{G^2}{\mathcal{Z}_b}
\sum_{m,n} e^{-\beta E_m}
 \langle n| {\cal O}_{\beta,-\vec k}(0) |m \rangle \langle m| {\cal O}_{\alpha,\vec k}(0) |n
 \rangle \, \delta(\omega-(E_m-E_n))
 \label{sigless}
\end{eqnarray}

\noindent where $\mathcal{Z}_b=\mathrm{Tr}\,e^{-\beta H_{\chi}}$ is
the equilibrium partition function of the ``bath'' and in the above
expressions the averages are solely with respect   to the bath
variables.  Upon relabelling $m \leftrightarrow n$ in the sum in the
definition (\ref{sigless}) and using the fact that these correlation
functions are  parity and rotational invariant\cite{kapusta} and
diagonal in the flavor basis we find the Kubo-Martin-Schwinger (KMS)
relation\cite{kapusta}
\begin{equation} \sigma^<_{ \alpha \beta}(k,\omega)  =
\sigma^>_{\alpha \beta}(k,-\omega) = e^{\beta \omega}
\sigma^>_{\alpha \beta}(k,\omega)\,. \label{KMS} \end{equation}
Using the spectral representation of the $\Theta(t-t')$ we find the
following representation for the retarded self-energy
\begin{equation} \Sigma^R_{\alpha \beta}(k;t-t')=
\int_{-\infty}^{\infty}\frac{dk_0}{2\pi} e^{ik_0(t-t')}
\widetilde{\Sigma}^R_{\alpha \beta}(k;k_0) \label{sigreta}
\end{equation} with \begin{equation}\label{sigofomega}
\widetilde{\Sigma}^R_{\alpha
\beta}(k;k_0)=\int_{-\infty}^{\infty}d\omega \frac{[\sigma^>_{\alpha
\beta} (k;\omega)-\sigma^<_{\alpha
\beta}(k;\omega)]}{\omega-k_0+i\epsilon}\,.
\end{equation} Using the condition (\ref{KMS}) the above spectral representation
can be written in a more useful manner as \begin{eqnarray}
\widetilde{\Sigma}^R_{\alpha \beta}(k;k_0) & = &
-\frac{1}{\pi}\int_{-\infty}^{\infty}d\omega
\frac{\mathrm{Im}\widetilde{\Sigma}^R_{\alpha
\beta}(k;\omega)}{\omega-k_0+i\epsilon}\,,\label{specsigret}\eea
where the imaginary part of the self-energy is given by \bea
\mathrm{Im}\widetilde{\Sigma}^R_{\alpha \beta}(k;\omega) & = & \pi
\sigma^>_{\alpha \beta}(k;\omega)\left[e^{\beta \omega}-1\right]
\label{imagpart}
\end{eqnarray} and is   positive for $\omega >0$.  Equation
(\ref{KMS}) entails that the imaginary part of the retarded
self-energy is an odd function of frequency, namely \begin{equation}
\mathrm{Im}\widetilde{\Sigma}^R_{\alpha \beta}(k; \omega) = -
\mathrm{Im}\widetilde{\Sigma}^R_{\alpha \beta}(k; -\omega)\; .
\label{odd}
\end{equation}
The relation (\ref{imagpart}) leads to the  following  results which
will  be useful later \be  \sigma^>_{\alpha \beta}(k;\omega)  =
\frac{1}{\pi}\mathrm{Im}\widetilde{\Sigma}^R_{\alpha
\beta}(k;\omega)\,n(\omega)~~;~~ \sigma^<_{\alpha \beta}(k;\omega) =
\frac{1}{\pi}\mathrm{Im}\widetilde{\Sigma}^R_{\alpha
\beta}(k;\omega)\,\left[1+n(\omega)\right]\label{sigs} \ee where
$n(\omega)= [e^{\beta \omega}-1]^{-1}$ is the Bose-Einstein
distribution function. Similarly from the definitions
(\ref{kernelkappa}) and (\ref{specrep}) and the condition
(\ref{KMS}) we find \begin{eqnarray} \mathcal{K}_{\alpha
\beta}(k;t-t') & = &\int_{-\infty}^{\infty}\frac{dk_0}{2\pi}
e^{ik_0(t-t')} \widetilde{\mathcal{K}}_{\alpha \beta}(k;k_0) \label{noisefou} \\
\widetilde{\mathcal{K}}_{\alpha \beta}(k;k_0)& = & \pi
\sigma^>_{\alpha \beta}(k;k_0)\left[e^{\beta k_0}+1\right]
\label{noisekernel}
\end{eqnarray}  whereupon using the condition (\ref{KMS}) leads to the
  generalized   fluctuation-dissipation relation \begin{equation}
\widetilde{\mathcal{K}}_{\alpha
\beta}(k;k_0)=\mathrm{Im}\widetilde{\Sigma}^R_{\alpha
\beta}(k;k_0)\coth\left[\frac{\beta
k_0}{2}\right]\,.\label{flucdiss}
\end{equation}

Thus we see that $\mathrm{Im}\widetilde{\Sigma}^R_{\alpha
\beta}(k;k_0)\,;\,\widetilde{\mathcal{K}}_{\alpha \beta}(k;k_0)$ are
odd and even functions of frequency respectively.

For the analysis below we  also need the following representation
(see eqn. (\ref{kernelsigma}))

\begin{equation}
\Sigma_{\alpha \beta}(k;t-t') = -i \int_{-\infty}^{\infty}
e^{i\omega(t-t')} \left[\sigma^>_{\alpha
\beta}(k;\omega)-\sigma^<_{\alpha \beta}(k;\omega)\right]d\omega =
\frac{i}{\pi}\int_{-\infty}^{\infty}
e^{i\omega(t-t')}\mathrm{Im}\widetilde{\Sigma}^R_{\alpha
\beta}(k;\omega) d\omega \label{sig}
\end{equation}

\noindent whose Laplace transform is given by

\begin{equation}
\widetilde{\Sigma}_{\alpha \beta}(k;s)\equiv \int^{\infty}_0 dt
e^{-st}\Sigma_{\alpha \beta}(k;t)= -\frac{1}{\pi}
\int^{\infty}_{-\infty}
\frac{\mathrm{Im}\widetilde{\Sigma}^R_{\alpha
\beta}(k;\omega)}{\omega+is}d\omega \label{laplasig}
\end{equation}

This spectral representation, combined with (\ref{specsigret}) lead
to the relation \begin{equation} \widetilde{\Sigma}^R_{\alpha
\beta}(k;k_0)=\widetilde{\Sigma}_{\alpha
\beta}(k;s=ik_0+\epsilon)\label{analyt}
\end{equation} The self energy and noise correlation kernels
$\widetilde{\Sigma},\widetilde{\mathcal{K}}$ are diagonal in the
flavor basis because the interaction is diagonal in this basis.
Namely, in the flavor basis

\be \widetilde{\Sigma}(k,\omega)= \Bigg(\begin{array}{cc}
                                                                   \widetilde{\Sigma}_{ee}(k,\omega) & 0 \\
                                                                   0 & \widetilde{\Sigma}_{\mu\mu}(k,\omega)
                                                                 \end{array}\Bigg)~~;~~\widetilde{\mathcal{K}}=
[1+2n(\omega)]\,\mathrm{Im}\widetilde{\Sigma}(k,\omega) =
\Bigg(\begin{array}{cc}
                                                                   \widetilde{\mathcal{K}}_{ee}(k,\omega) & 0 \\
                                                                   0 &\widetilde{\mathcal{K}}_{\mu\mu}(k,\omega)
                                                                 \end{array}\Bigg) \,.
                                                                 \label{flavorsigK}\ee

 In the\emph{ mass} basis these kernels are given by \be
\widetilde{\Sigma}=
\frac{1}{2}\Big(\widetilde{\Sigma}_{ee}+\widetilde{\Sigma}_{\mu\mu}\Big)\,\mathds{1}
 +\frac{1}{2}\Big(\widetilde{\Sigma}_{ee}-\widetilde{\Sigma}_{\mu\mu}\Big)  \,\Bigg(\begin{array}{cc}
                                                                   \cos 2\theta & \sin 2\theta \\
                                                                   \sin 2\theta &
                                                                   -\cos
                                                                   2\theta
                                                                 \end{array}\Bigg)
                                                                 \label{Sigmamass}\ee and

\be \widetilde{\mathcal{K}}=
\frac{1}{2}\Big(\widetilde{\mathcal{K}}_{ee}+\widetilde{\mathcal{K}}_{\mu\mu}\Big)\,\mathds{1}
 +\frac{1}{2}\Big(\widetilde{\mathcal{K}}_{ee}-\widetilde{\mathcal{K}}_{\mu\mu}\Big)  \,\Bigg(\begin{array}{cc}
                                                                   \cos 2\theta & \sin 2\theta \\
                                                                   \sin 2\theta &
                                                                   -\cos
                                                                   2\theta
                                                                 \end{array}\Bigg)\,.
                                                                 \label{kapamass}\ee

\section{Dynamics: solving the Langevin
equation}\label{sec:langevin}

The solution of the equation of motion (\ref{langevin}) can be found
by Laplace transform. Define the Laplace transforms

\be \label{laplapsi} \widetilde{\Psi}_{\alpha,\vk}( s)  =
\int^{\infty}_0 dt e^{-st}\Psi_{\alpha;\vec k}(t)
~~;~~\widetilde{\xi}_{\alpha,\vec k}(s)  = \int^{\infty}_0 dt
e^{-st}\xi_{\alpha,\vec k}(t) \ee along with the Laplace transform
of the self-energy given by eqn. (\ref{laplasig}). The Langevin
equation in Laplace variable becomes the following algebraic matrix
equation

\begin{equation}\label{langeqlapla}
\Bigg[(s^2+k^2)\delta_{\alpha \beta}+\mathbb{M}^2_{\alpha
\beta}+\mathds{V}_{\alpha \beta}+\widetilde{\Sigma}_{\alpha
\beta}(k;s) \Bigg]\widetilde{\Psi}_{\beta,\vec k}(s)=
\Pi_{0,\alpha,\vec k}+s\,\Psi_{0,\alpha,\vec
k}+\widetilde{\xi}_{\alpha,\vec k}(s)
\end{equation}
\noindent where we have used the initial conditions (\ref{bcfin}).
The solution in real time can be written in a more compact manner as
follows. Introduce the matrix function \be \label{Gfunc}
\widetilde{G}(k;s)=
\Bigg[(s^2+k^2)\mathds{1}+\mathbb{M}^2+\mathds{V}
+\widetilde{\Sigma}(k;s) \Bigg]^{-1} \ee and its anti-Laplace
transform \be G_{\alpha \beta}(k;t) = \int_\mathcal{C}
\frac{ds}{2\pi i}\,\widetilde{G}_{\alpha
\beta}(k;s)\,e^{st}\label{Gfunlap} \ee where $\mathcal{C}$ refers to
the Bromwich contour, parallel to the imaginary axis in the complex
$s$ plane to the right of all the singularities of
$\widetilde{G}(k;s)$. This   function obeys the initial conditions
\be G_{\alpha \beta}(k;0)=0 ~~;~~ \dot{G}_{\alpha \beta}(k;0)=1\,.
\label{inicondG}\ee In terms of this auxiliary function the solution
of the Langevin equation (\ref{langevin}) in real time is given by
\begin{equation}
\Psi_{\alpha,\vk}(t;\Psi^0;\Pi^0;\xi) = \dot{G}_{\alpha
\beta}(k;t)\,\Psi^0_{\beta\vec k} + {G}_{\alpha \beta}(k;t)\,
 \Pi^0_{\beta,\vec k} + \int^t_0
G_{\alpha \beta}(k;t')~\xi_{\beta,\vec k}(t-t') dt' \,,
\label{inhosolution}
\end{equation} where the dot stands for derivative with respect to
time. In the flavor basis we find
 \be
\widetilde{G}_f(k;s)= \mathcal{S}(k;s)\
\Bigg[\Big(s^2+\overline{\omega}^2(k)+\overline{\Sigma}(k;s)\Big)\mathds{1}
+ \frac{\delta\,M^2}{2}\,\Bigg(\begin{array}{cc}
                   \cos2\theta-\Delta(k;s)  & -\sin 2\theta \\
                        -\sin 2\theta& - \cos 2\theta +\Delta(k;s) \\
                       \end{array}\Bigg) \Bigg] \label{Gflavor}\ee whereas in the mass basis we find

\be \widetilde{G}_m(k;s)= \mathcal{S}(k;s)\ \Bigg[\Big(s^2+
\overline{\omega}^2(k)+\overline{\Sigma}(k;s) \Big) \mathds{1}+
\frac{\delta\,M^2}{2}\,\Bigg(\begin{array}{cc}
                    1-\Delta(k;s)\,\cos2\theta    &  \Delta(k;s)\,\sin 2\theta \\
                        \Delta(k;s)\,\sin 2\theta& -  1+\Delta(k;s)\,\cos2\theta     \\
                       \end{array}\Bigg) \Bigg] \label{Gmass}\ee
where $\overline{M}^2$ and $\delta M^2$ were defined by eqn.
(\ref{MbarDelM}) and  to simplify notation we defined \bea
\overline{\omega}(k) & = &
\sqrt{k^2+\overline{M}^2} \label{overomega}\\
\overline{\Sigma}(k;s)  & = &
\frac{1}{2}\Big(\widetilde{\Sigma}_{ee}(k;s)+V_{ee}+\widetilde{\Sigma}_{\mu\mu}(k;s)+V_{\mu\mu}\Big)\label{oversig}\\
\Delta(k;s) & = &
\frac{\widetilde{\Sigma}_{ee}(k;s)+V_{ee}-\widetilde{\Sigma}_{\mu\mu}(k;s)-V_{\mu\mu}}{M^2_2-M^2_1}
\label{Delta} \eea and

\be \mathcal{S}(k;s) =   \Bigg[\Big(s^2 +
\overline{\omega}^{\,2}(k) +\overline{\Sigma}(k;s)\Big)^2 -
\Big(\frac{\delta\, M^2}{2}\Big)^2 \,\Big[ (\cos 2\theta -
\Delta(k;s))^2+(\sin 2\theta)^2 \Big]\Bigg]^{-1} \ee

In what follows we define the analytic continuation of the
quantities defined above with the same nomenclature to avoid
introducing further notation, namely

\be  \overline{\Sigma}(k;\omega)    \equiv
\overline{\Sigma}(k;s=i\omega+\epsilon) ~~;~~ \Delta(k;\omega)
\equiv   \Delta(k;s=i\omega+\epsilon)\,. \label{overs}\ee  Their
real and imaginary parts are given by \bea
\overline{\Sigma}_R(k;\omega) & = &
\frac{1}{2}\left[\Sigma_{R,ee}(k,\omega)+\Sigma_{R,\mu\mu}(k,\omega)+V_{ee}+V_{\mu\mu}\right]
\label{sigre} \\ \overline{\Sigma}_I(k;\omega) & = &
\frac{1}{2}\left[\Sigma_{I,ee}(k,\omega)+\Sigma_{I,\mu\mu}(k,\omega)\right]
\label{sigim}\\\Delta_R(k;\omega) & = &
\frac{1}{\delta\,M^2}\left[\Sigma_{R,ee}(k,\omega)-\Sigma_{R,\mu\mu}(k,\omega)+V_{ee}-V_{\mu\mu}\right]\\
\Delta_I(k;\omega) & = &
\frac{1}{\delta\,M^2}\left[\Sigma_{I,ee}(k,\omega)-\Sigma_{I,\mu\mu}(k,\omega)
\right]\eea We remark that while the matter potential $V$ is of   of
order $G$, $\Sigma$ is of   order $G^2$. Therefore, in perturbation
theory \be \overline{\Sigma}_R(k;\omega) >>
\overline{\Sigma}_{I}(k;\omega)~~;~~\Delta_R(k;\omega)>>\Delta_I(k;\omega)\,.
\label{greatt} \ee This inequality also holds in the standard model,
where the matter potential is of order $G_F$\cite{notzold} while the
absorptive part that determines the relaxation rates is of order
$G^2_F$. This perturbative inequality will be used repeatedly in the
analysis that follows, and we emphasize that it holds in the correct
description of neutrinos propagating in a medium.

\section{ Single particles and quasiparticles}\label{sec:quasi}

Exact single particle states are determined by the position of the
isolated poles of the Green's function in the complex $s-$ plane.
Before we study the interacting case, it proves illuminating to
first study the \emph{free}, non-interacting case.

\vspace{2mm}

\subsection{Free   case: $G=0$}

 To begin the analysis, an example helps to clarify this
formulation: consider the non-interacting case $G=0$ in which
$\overline{\Sigma} =0;\Delta =0 $. In this case
$\widetilde{G}_{f,m}(k;s)$ have simple poles at $s = \pm i
\omega_1(k)$ and $\pm i\omega_2(k)$ where \be \omega_i(k) =
\sqrt{k^2+M^2_i}~~;~~i=1,2 \label{freqs0}\,. \ee Computing the
residues at these simple poles we find in the flavor basis \be
G_f(k;t) = \frac{\sin (\omega_1(k)t)}{\omega_1(k)}\,
\mathds{R}^{(1)} (\theta)
                                                       + \frac{\sin
                                                       (\omega_2(k)t)}{\omega_2(k)}
                                                       \, \mathds{R}^{(2)}
                                                       (\theta)
                                                       \label{Gft}\ee
where   we have introduced the   matrices \be    \mathds{R}^{(1)}
(\theta)   = \Bigg(\begin{array}{cc}
                                                         \cos^2\theta & -\cos\theta \sin \theta \\
                                                         -\cos\theta \sin \theta & \sin^2\theta \\
                                                       \end{array}\Bigg)
                                                       = U(\theta)\, \Bigg(\begin{array}{cc}
                                                         1 & 0 \\
                                                         0 & 0 \\
                                                       \end{array}\Bigg)\,  U^{-1}(\theta)  \label{R1} \ee
                                                       \be  \mathds{R}^{(2)}
                                                       (\theta)    =    \Bigg(\begin{array}{cc}
                                                         \sin^2\theta &  \cos\theta \sin \theta \\
                                                          \cos\theta \sin \theta & \cos^2\theta \\
                                                       \end{array}\Bigg) = U(\theta)\, \Bigg(\begin{array}{cc}
                                                         0 & 0 \\
                                                         0 & 1 \\
                                                       \end{array}\Bigg)\,  U^{-1}(\theta)
                                                       \label{R2}\ee

In the mass basis
 \be G_m(k;t) = \Bigg(\begin{array}{cc}
                  \frac{\sin (\omega_1(k)t)}{\omega_1(k)} & 0 \\
                  0 & \frac{\sin (\omega_2(k)t)}{\omega_2(k)}
                \end{array} \Bigg) \label{Gmt}\ee \noindent with the
                relation

                \be G_f(k;t) =  U(\theta)\, G_m(k;t)\,U^{-1}(\theta)
                \label{unitrafo}\ee and $U(\theta)$ is given by
                \ref{trafo}. Consider for simplicity an initial condition with $\Psi^0\neq
0;\Pi^0 =0$ in both cases, flavor and mass. The expectation value of
the flavor fields $\Phi_\alpha$ in the reduced density matrix
(\ref{totave}) is given by

\be   \Big\langle\Big(\begin{array}{c}
        \Psi_{e,\vk}(t) \\
              \Psi_{\mu,\vk}(t) \\
            \end{array}\Big) \Big\rangle  =  \Bigg[ {\cos
            (\omega_1(k)t)} \,\mathds{R}^{(1)} (\theta)
                                                       +  {\cos (\omega_2(k)t)}\,\mathds{R}^{(2)} (\theta)\Bigg] \Bigg(\begin{array}{c}
        \Psi^0_{e,\vk} \\
              \Psi^0_{\mu,\vk} \\
            \end{array}\Bigg)\label{psioft}            \ee and
that   for the fields in the mass basis is \be  \Big\langle
\Big(\begin{array}{c}
        \Psi_1(k;t) \\
              \Psi_2(k;t) \\
            \end{array}\Big) \Big\rangle  = \Bigg(\begin{array}{c}
        \Psi^0_{1,\vk}\,{\cos
            (\omega_1(k)t)} \\
              \Psi^0_{2,\vk}\,{\cos (\omega_2(k)t)} \\
            \end{array}\Bigg) \ee  These are precisely the solutions of the classical equations of motion
 in terms of flavor and mass eigenstates, namely  \bea \phi_e(k;t)  & = &  \cos\theta\, \varphi_1(k;t)+ \sin \theta
 \,
 \varphi_2(k;t) \nonumber\\ \phi_\mu(k;t) & = &
 \cos\theta \, \varphi_2(k;t) - \sin \theta \, \varphi_2(k;t) \eea where
\be \varphi_1(k;t) = \varphi_1(k;0)\,\cos\omega_1(k)t ~~;~~
\varphi_2(k;t) = \varphi_2(k;0)\,\cos\omega_2(k)t \label{fioft}\ee
for vanishing initial canonical momentum and the initial values are
given in terms of flavor fields   by \bea \varphi_1(k;0)  & =  &
\cos\theta \, \phi_e(k;0)-\sin\theta \,
\phi_\mu(k;0)\nonumber\\
\varphi_2(k;0)  & =  & \cos\theta \,\phi_\mu(k;0)+\sin\theta \,
\phi_e(k;0) \label{inifis}\eea inserting (\ref{fioft}) with the
initial conditions (\ref{inifis}) one recognizes that the solution
(\ref{psioft}) is the expectation value of the classical equation of
motion with initial conditions on the flavor fields and vanishing
initial  canonical momentum.

It is  convenient to separate the positive  (particles) and negative
(antiparticles) frequency components by considering an initial
condition with $\Pi^0_{\alpha,\vec{k}} \neq 0$,  in such a way that
the time dependence is determined by phases corresponding to the
propagation of particles (or antiparticles). Without mixing
($\theta=0$) this is achieved by choosing the following initial
conditions \be \Pi^0_{\alpha,\vec{k}} = \mp i \Omega_\alpha(k)  \,
\Psi^0_{\alpha,\vec{k}}\label{inipart} \ee for particles ($-$) and
antiparticles ($+$) respectively, as in eq. (\ref{aadaggerflav}).
This choice of initial conditions leads to  the result \be \langle
\langle \Psi_{\alpha,\vec{k}}(t) \rangle \rangle = \Bigg\{
\mathds{R}^{(1)}_{\alpha\beta}(\theta) \left[\cos(\omega_1(k)t)\mp
i\frac{\Omega_\beta(k)}{\omega_1(k)}\,\sin(\omega_1(k)t)\right]+
 \mathds{R}^{(2)}_{\alpha\beta}(\theta)
\left[\cos(\omega_2(k)t)\mp
i\frac{\Omega_\beta(k)}{\omega_2(k)}\,\sin(\omega_2(k)t)\right]\Bigg\}\Psi^0_{\beta,\vec{k}}
\label{evolu} \ee

It is clear from (\ref{evolu}) that no single choice of frequencies
$\Omega_\beta(k)$ can lead uniquely to time evolution in terms of
single particle/antiparticle phases $e^{\mp i \omega_{1,2}(k)t}$.
This is a consequence of the ambiguity in the definition of flavor
states as discussed in detail in section (\ref{flavmass}). However,
for the cases in which   $|M^2_1-M^2_2|\ll (k^2+\overline{M}^2) $ ,
relevant for relativistic mixed neutrinos, and for
$K^0\overline{K}^0$ and $B^0\overline{B}^0$ mixing, the positive and
negative frequency components can be \emph{approximately} projected
out as follows. Define

\be \overline{\omega}(k)  = \sqrt{k^2+\overline{M}^2}
\label{aveomega}\ee in the nearly degenerate or relativistic regime
when $|\delta M^2|/\,\overline{\omega}^{\, 2}(k)  \ll 1$

\be \omega_1(k) =  \overline{\omega}(k)\left[1 - \frac{\delta
M^2}{4\, \overline{\omega}^{\, 2}(k) }
+\mathcal{O}\Bigg(\frac{\delta M^2}{\overline{\omega}^{\, 2}(k)
}\Bigg)^2\right]~~;~~\omega_2(k) =\overline{\omega}(k)\left[1 +
\frac{\delta M^2}{4\, {\overline{\omega}^{\,2 }(k)} }
+\mathcal{O}\Bigg(\frac{ \delta M^2}{\overline{\omega}^{\,2 }(k)
}\Bigg)^2\right]\,. \label{avedel}\ee

Taking $\Omega_\beta(k) = \overline{\omega}(k) $ and choosing for
example the negative sign (positive frequency component) in
(\ref{evolu}) we find

\be   \langle \Psi_{\alpha,\vec{k}}(t) \rangle   = \Bigg\{
\mathds{R}^{(1)}_{\alpha\beta}(\theta) \,e^{-i\omega_1(k)t}+
 \mathds{R}^{(2)}_{\alpha\beta}(\theta)\,e^{-i\omega_2(k)t} +\mathcal{O}\Bigg(\frac{ \delta
M^2}{\overline{\omega}^{\, 2 }(k)}\Bigg)
 \Bigg\}\Psi^0_{\beta,\vec{k}}  \label{evoluphase} \ee

Consider the following initial condition \be \Psi^0_{\vec{k}} =
\left(
                         \begin{array}{c}
                           \Psi^0_{e,\vec{k}} \\
                           0 \\
                         \end{array}
                       \right) \label{iniPsi}
 \ee neglecting the corrections in (\ref{evoluphase}) we find
 \be |\langle   \Psi_{\mu,\vec{k}}(t) \rangle   |^2 =
\sin^2 2\theta \, \sin^2\Bigg(
 \frac{\delta M^2}{4\,\overline{\omega}(k)}\,
 t\Bigg) \Psi^0_{e,\vec{k}}+ \mathcal{O}\Bigg(\frac{ \delta
M^2}{\overline{\omega}^{\,2 }(k)}\Bigg)^2
 \label{proba}\ee which is identified  with  the usual result for the oscillation transition
 probability $\Psi_e \rightarrow \Psi_\mu$ upon neglecting the corrections.
 \subsection{Interacting theory, $G\neq 0$}

For $G\neq 0$,  the self-energy as a function of frequency and
momentum is in general complex, the imaginary part arises from
multiparticle thresholds. When the imaginary part of the self-energy
does not vanish at the value of the frequency corresponding to the
dispersion relation of the free particle states, these particles can
decay and no longer appear as asymptotic states. The poles in the
Green's function move off the physical sheet into a higher Riemann
sheet, the particles now become resonances.

\begin{figure}[ht!]
\begin{center}
\includegraphics[height=2in,width=3in,keepaspectratio=true]{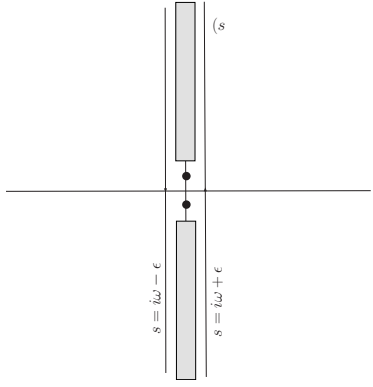}
\caption{Bromwich contour in s-plane, the shaded region denotes the
cut discontinuity from multiparticle thresholds  across the
imaginary axis, the filled circles represent the single particle
poles. } \label{fig:contoursplanepoles}
\end{center}
\end{figure}

Single particle states correspond to true poles of the propagator
(Green's function) in the physical sheet, which are necessarily away
from the multiparticle thresholds. This case is depicted in fig.
(\ref{fig:contoursplanepoles}).

Let us consider the Green's function in the flavor basis eqn.
(\ref{Gflavor}). The single particle poles are determined by the
poles of $\mathcal{S}(k;s)$ on the imaginary axis away from the
multiparticle cuts. These are determined by the roots of the
following equations

\bea \Omega^2_1(k) & = & \overline{\omega}^{\,2}(k)
+\overline{\Sigma}_R(k;\Omega_1(k)) - \frac{\delta M^2}{2}
\Big[\big(\cos 2\theta - \Delta_R(k;\Omega_1(k))\big)^2+\big(\sin
2\theta \big)^2\Big]^{\frac{1}{2}} \label{Omega1}
\\
\Omega^2_2(k) & = & \overline{\omega}^{\,2}(k)  +
\overline{\Sigma}_R(k;\Omega_2(k)) + \frac{\delta M^2}{2}
 \Big[\big(\cos 2\theta - \Delta_R(k;\Omega_2(k))\big)^2+\big(\sin
2\theta \big)^2\Big]^{\frac{1}{2}}  \label{Omega2} \eea along with
the conditions \be \overline{\Sigma}_I(k;\Omega_{1,2}(k))=0 ~~;~~
\Delta_I(k;\Omega_{1,2}(k))=0 \ee where the subscripts $R,I$ refer
to the real and imaginary parts respectively. Evaluating the
residues at the single particle poles and using that the real and
imaginary parts of the self-energies are even and odd functions of
frequency respectively, we find

\be  G_{f}(k;t) =   Z^{(1)}_k \,\frac{\sin
(\Omega_1(k)t)}{\Omega_1(k)} ~~
\mathds{R}^{(1)}\big(\theta^{(1)}_m(k)\big)+ Z^{(2)}_k \,\frac{\sin
(\Omega_2(k)t)}{\Omega_2(k)}~~
\mathds{R}^{(2)}\big(\theta^{(2)}_m(k)\big) +  G_{f,cut}(t)
\label{Gft2}\ee  where $G_{f,cut}(t)$ is the contribution from the
multiparticle cut, the matrices $\mathds{R}^{(1,2)}$ are given by
(\ref{R1},\ref{R2})  and $\theta^{1,2}_m(k)$ are the mixing angles
\emph{in the medium}

\be \cos2\theta^{i}_m(k) = \frac{\cos2\theta -
\Delta_R(\Omega_{i}(k))}{\Big[\big(\cos 2\theta -
\Delta_R(k;\Omega_{i}(k))\big)^2+\big(\sin 2\theta
\big)^2\Big]^{\frac{1}{2}}} ~~;~~ \sin2\theta^{i}_m(k) =
\frac{\sin2\theta}{\Big[\big(\cos 2\theta -
\Delta_R(k;\Omega_{i}(k))\big)^2+\big(\sin 2\theta
\big)^2\Big]^{\frac{1}{2}}} \,.\label{tetamed} \ee for $i=1,2$.
 The wave function renormalization constants are given by

 \be Z^{(i)}_k =
 \Bigg[1-\frac{1}{2\omega}\Big(\overline{\Sigma}^{'}_R(k;\omega) - (-1)^i
 \frac{\delta M^2}{2} \cos2\theta^{i}_m(k)
 \Delta^{'}_R(k;\omega)\Big)\Bigg]^{-1}_{\omega=\Omega^{i}(k)} \ee
 where the prime stands for derivative with respect to $\omega$.
At asymptotically long time the contribution from the cut
$G_{f,cut}(t) \sim t^{-\alpha}$ where $\alpha$ is determined by the
behavior of the self-energy at threshold\cite{threshold,maiani}.

In perturbation theory and in the   limit  $
\overline{\omega}(k)^{\,2} \gg |\delta M^2|$ the dispersion
relations (\ref{Omega1},\ref{Omega2}) can be solved by writing

\be \pm \Omega^i(k) = \pm (\overline{\omega}(k)
+\delta\omega_i(k))\,, \label{Omegs} \ee   we find \be
\label{delomegas} \delta \omega_i(k) =
\frac{\overline{\Sigma}_R(k;\overline{\omega}(k))}{2\,\overline{\omega
}(k)} + (-1)^i ~~ \frac{\delta M^2}{4\, \overline{\omega} (k)}
\,\overline{\varrho}(k) \ee where we defined \be \varrho(k;\omega) =
\Bigg[(\cos 2\theta -
\Delta_R(k;\omega))^2+(\sin2\theta)^2\Bigg]^{\frac{1}{2}}
\label{varrho}\ee and the shorthand \be\overline{\varrho}(k)   =
\varrho(k;\omega=\overline{\omega}(k)) \label{varrhok}  \,.   \ee To
leading order in the perturbative expansion and in $\delta
M^2/\overline{\omega}^{\,2}(k) $ we find $\theta^{(1)}_m(k) =
\theta^{(2)}_m(k)= \theta_m(k)$. Gathering these results, we find
the dispersion relations and mixing angles in the medium to be given
by the following relations, \bea \Omega_1(k) & = &
\overline{\omega}(k) +
\frac{\overline{\Sigma}_R(k;\overline{\omega}(k))}{2\,\overline{\omega
}(k)} - \frac{\delta M^2}{4\, \overline{\omega} (k)}
\,\overline{\varrho}(k) \label{omegone}\\
\Omega_2(k) & = & \overline{\omega}(k) +
\frac{\overline{\Sigma}_R(k;\overline{\omega}(k))}{2\,\overline{\omega
}(k)} +   \frac{\delta M^2}{4\, \overline{\omega} (k)}
\,\overline{\varrho}(k)\,,\label{omegtwo}\eea and \be
\cos2\theta_m(k) = \frac{\cos2\theta -
\Delta_R(k;\overline{\omega}(k))}{\Big[\big(\cos 2\theta -
\Delta_R(k;\overline{\omega}(k))\big)^2+\big(\sin 2\theta
\big)^2\Big]^{\frac{1}{2}}} ~~;~~ \sin2\theta_m(k) =
\frac{\sin2\theta}{\Big[\big(\cos 2\theta -
\Delta_R(k;\overline{\omega}(k))\big)^2+\big(\sin 2\theta
\big)^2\Big]^{\frac{1}{2}}} \,.\label{tetamedav} \ee These
dispersion relations and mixing angles have exactly the \emph{same
form} as those obtained in the field theoretical studies of neutrino
mixing in a medium\cite{notzold,boyho}.

\subsection{Quasiparticles and relaxation. }

\begin{figure}[ht!]
\begin{center}
\includegraphics[height=2in,width=4in,keepaspectratio=true]{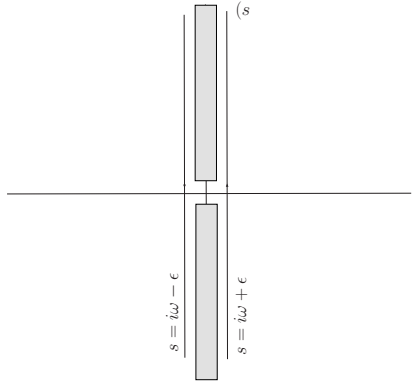}
\caption{Bromwich contour in s-plane, the shaded region denotes the
cut discontinuity from multiparticle thresholds  across the
imaginary axis. } \label{fig:contoursplane}
\end{center}
\end{figure}

Even a particle that is stable in the vacuum acquires a width in the
medium through several processes, such as collisional broadening or
Landau damping\cite{kapusta}. In this case there are no isolated
poles in the Green's function in the physical sheet, the poles move
off the imaginary axis in the complex $s-$plane on to a second or
higher Riemann sheet. The Green's function now features branch cut
discontinuities across the imaginary axis perhaps with isolated
regions of analyticity. The inverse Laplace transform is now carried
out by wrapping around the imaginary axis as shown in fig.
(\ref{fig:contoursplane}), and the real time Green's function is
given by \be G_{\alpha\beta}(k;t) = \frac{i}{\pi}
\int_{-\infty}^\infty d\omega \,e^{i\omega t} ~~
\mathrm{Im}\widetilde{G}_{\alpha\beta}(k;s=i\omega +
\epsilon)\label{goft} \ee Under the validity of perturbation theory,
when the inequality (\ref{greatt}) is fulfilled we consistently keep
terms up to $\mathcal{O}(G^2)$ and  find the imaginary part to be
given by the following expression \be \label{ImG}
\mathrm{Im}\widetilde{G}(k;s=i\omega + \epsilon) =
\frac{-\mathds{A}(D_- \gamma_+ +D_+ \gamma_-)+\mathds{B}(D_+ D_-
-\gamma_+\gamma_-)}{(D^2_+ +\gamma^2_+)(D^2_- +\gamma^2_-)}\ee where
we have introduced   \bea D_{\pm}(k;\omega)  & = &
-\omega^2+\overline{\omega}^{\,2}_k +
\overline{\Sigma}_R(k;\omega)~\mp \frac{1}{2}\delta M^2
\varrho(k;\omega)  \label{Dpm} \\
\gamma_\pm(k;\omega) & = &
 \frac{1}{2}\left(1\pm\cos2\theta_m(\omega,k)\right)\,\Sigma_{I,ee}(k;\omega)+
 \frac{1}{2}\left(1\mp\cos2\theta_m(\omega,k)\right) \, \Sigma_{I,\mu\mu}(k;\omega) \label{gammapm}
\eea $\Sigma_{R,I}$ are the real and imaginary parts of the self
energy respectively, with  \be \Delta_{R}(k;\omega)   =
\frac{1}{\delta
M^2}\left[\Sigma_{R,ee}(k;\omega)+V_{ee}-\Sigma_{R,\mu\mu}(k;\omega)-V_{\mu\mu}\right]
\label{deltaR}\,, \ee and \be   \mathds{A}(k;\omega)    =   \left[
-\omega^2+\overline{\omega}^{\,2}_k +\overline{\Sigma}_R(k;\omega)
\mathds{1} \right] +   \frac{\delta M^2}{2} \left( \begin{array}{cc}
                                                                        \cos2\theta - \Delta_R(k;\omega) & -\sin2\theta \\
                                                                         -\sin2\theta & -\cos2\theta +\Delta_R(k;\omega)
                                                                      \end{array}
                                                                      \right)
                                                                      \;,
                                                                      \label{matrixA}\ee

\be \mathds{B}(k;\omega) = \left(
                             \begin{array}{cc}
                               \Sigma_{I,\mu\mu}(k;\omega) & 0 \\
                               0 & \Sigma_{I,ee}(k;\omega) \\
                             \end{array}
                           \right)\,. \label{matrixB}\ee The denominator in (\ref{ImG}) features complex zeroes for
\be D_\pm(k;\omega)+\gamma_\pm(k;\omega) = 0\,. \label{poles} \ee

Near these zeroes $\mathrm{Im}\widetilde{G}(k;s=i\omega + \epsilon)$
has the typical Breit-Wigner form for resonances. The dynamical
evolution at long times is dominated by the complex poles in the
upper half $\omega-$ plane associated with these resonances. In
perturbation theory the complex poles of
$\mathrm{Im}\widetilde{G}(k;s=i\omega + \epsilon)$ occur at

\be \omega = \pm \Omega_1(k) +i \frac{\Gamma_1(k)}{2}\label{pole1}
\ee and \be \omega = \pm \Omega_2(k) +i
\frac{\Gamma_2(k)}{2}\label{pole2} \ee where $\Omega_{1,2}(k)$  are
given by (\ref{Omega1},\ref{Omega2}) and

\be \frac{\Gamma_{1}(k)}{2} =
\frac{\gamma_+(k,\Omega_1(k))}{2\,\Omega_1(k)} \approx
\frac{\gamma_+(k,\overline{\omega}(k))}{2\,\overline{ \omega}(k)}
~~;~~ \frac{\Gamma_{2}(k)}{2} =
\frac{\gamma_-(k,\Omega_2(k))}{2\,\Omega_2(k)} \approx
\frac{\gamma_-(k,\overline{\omega}(k))}{2\,\overline{\omega}(k)}\label{Gamas}\ee

These relaxation rates can be written in an  illuminating manner

\bea \Gamma_1(k)  & = &   \Gamma_{ee}(k)
\cos^2\theta_m(k)+\Gamma_{\mu\mu}(k)\sin^2\theta_m(k)
\label{Gama1ee} \\\Gamma_2(k)  & = &   \Gamma_{\mu\mu}(k)
\cos^2\theta_m(k)+\Gamma_{ee}(k)\sin^2\theta_m(k)
\label{Gama2mumu}\eea where \be\Gamma_{\alpha\alpha}(k) =
\frac{\Sigma_{I,\alpha\alpha}(k;\ovom)}{ \ovom}\ee are the
relaxation rates of the flavor fields in \emph{absence} of mixing.
These relaxation rates are similar to those proposed within the
context of flavor conversions in supernovae\cite{raffsigl}, or
active-sterile oscillations\cite{aba,foot,dibari}.

We carry out the frequency integral in (\ref{goft}) by approximating
the integrand as a sum of two Breit-Wigner Lorentzians near $\omega
= \pm \Omega_{1,2}(k)$ with the following result in the flavor
basis,

\bea  G_{f}(k;t) = && Z^{(1)}_k \,e^{-\frac{\Gamma_1(k)}{2}t}
\,\Bigg[\frac{\sin(\Omega_1(k)t)}{
{\Omega_1(k)}}\,\mathds{R}^{(1)}(\theta_m(k))-
\frac{\widetilde{\gamma}(k)}{2}\,\frac{\cos(\Omega_1(k)t)}{\Omega_1(k)}\,\mathds{R}^{(3)}(\theta_m(k))\Bigg]+
\nonumber \\ &&Z^{(2)}_k\, e^{-\frac{\Gamma_2(k)}{2}t}\,\Bigg[
\frac{\sin(\Omega_2(k)t)}{\Omega_2(k)}\,\mathds{R}^{(2)}(\theta_m(k))+
\frac{\widetilde{\gamma}(k)}{2}\,\frac{\cos(\Omega_2(k)t)}{\Omega_2(k)}\,\mathds{R}^{(3)}(\theta_m(k))\Bigg]
                                                       \label{Gdecay}
                                                       \eea where
again we have used the approximation $|\delta M^2|\ll
\overline{\omega}^{\,2}(k) $ and introduced

\be \label{R3} \mathds{R}^{(3)}(\theta ) = \sin 2\theta ~~ \left(
                                             \begin{array}{cc}
                                               \sin 2\theta  & \cos 2\theta  \\
                                               \cos 2\theta  & -\sin 2\theta  \\
                                             \end{array}
                                           \right) = \sin 2\theta \, U(\theta )~\left(
                                             \begin{array}{cc}
                                               0 & 1 \\
                                               1 & 0 \\
                                             \end{array}
                                           \right)~U^{-1}(\theta ) \ee and \be \label{gamatil}
                                           \widetilde{\gamma}(k)=
                                           \frac{\Sigma_{I,ee}(k;\overline{\omega}(k) -\Sigma_{I,\mu\mu}(k;\overline{\omega}(k) )}{\delta
                                           M^2\,\overline{\varrho}(k)}\,.  \ee

Under the assumption that $\Sigma_{R,\alpha,\alpha} \gg
\Sigma_{I,\alpha,\alpha}$ it follows that $\widetilde{\gamma}(k) \ll
1$. As in the previous section, we can approximately project the
positive frequency component by choosing the initial condition
(\ref{inipart}) with $\Omega_\alpha = \overline{\omega}(k)$, leading
to the result

\bea \label{posires}  \langle \Psi_{\alpha,\vec{k}}(t) \rangle =
e^{-i W(k)t-\frac{\overline{\Gamma}(k)}{2}t}  &&\Bigg\{ Z^{(1)}_k
\,e^{ i\Delta\omega(k)t-\frac{\Delta \Gamma(k)}{2}t}~\Big(
\mathds{R}^{(1)} (\theta_m(k))  + i\frac{\widetilde{\gamma}(k)}{2}\,
\mathds{R}^{(3)} (\theta_m(k)) \Big) + \nonumber \\ && Z^{(2)}_k
\,e^{- i\Delta\omega(k)t+\frac{\Delta \Gamma(k)}{2}t}~\Big(
 \mathds{R}^{(2)}_{\alpha,\beta}(\theta) - i\frac{\widetilde{\gamma}(k)}{2}\, \mathds{R}^{(3)}
(\theta_m(k)) \Big) +\mathcal{O}\Bigg(\frac{ \delta
M^2}{\overline{\omega}^{\, 2 }(k) }\Bigg)
 \Bigg\}\Psi^0_{\beta,\vec{k}}  \eea where

 \bea W(k) & = & \overline{\omega}(k)
 +\frac{\overline{\Sigma}_R(k;\overline{\omega}(k))}{4\,\overline{\omega}(k)}
 \label{bigOm} \\ \frac{\overline{\Gamma}(k)}{2} & = &
 \frac{1}{4\overline{\omega}(k)}\,\Big[\Sigma_{I,ee}(k,\overline{\omega}(k)+\Sigma_{I,\mu\mu}(k,\overline{\omega}(k)
 \Big]\label{medGa} \\
  \Delta \omega(k) & =  & \frac{\delta
 M^2\,\overline{\varrho}(k)}{4\,\overline{\omega}(k)} \label{deltome}\\ \frac{\Delta\Gamma(k)}{2} & = &   \frac{\cos2\theta_m}{4~\overline{\omega}(k)}
 \,\Big[\Sigma_{I,ee}(k,\overline{\omega}(k)-\Sigma_{I,\mu\mu}(k,\overline{\omega}(k)
 \Big] \label{deltom}\eea

With the initial condition (\ref{iniPsi}) we now find the long time
evolution of the transition probability $\Psi_e \rightarrow
\Psi_\mu$ \be |  \langle \Psi_{\mu,\vec{k}}(t) \rangle   |^2 \sim
\frac{\sin^2 2\theta_m(k)}{4}  \Big[e^{-   {\Gamma_1(k)} t}+ e^{-
 {\Gamma_2(k)}  t}-2~e^{-\frac{1}{2}(\Gamma_1(k)+\Gamma_2(k))
t}\cos\big(2\Delta \omega(k) t\big) \Big]~\Psi^0_{e,\vec{k}}\ee
where we have neglected perturbatively small terms by setting
$Z^{(i)}\sim 1 ~;~ \widetilde{\gamma}(k)\sim 0$ in prefactors. The
solution (\ref{posires}) can be written in the following
illuminating form \be  \label{posires2}  \langle
\Psi_{\alpha,\vec{k}}(t) \rangle
  = e^{-i W(k)t-\frac{\overline{\Gamma}(k)}{2}t}\,
\mathcal{U}\big(\theta_m(k)\big)~ \Bigg(
                                \begin{array}{cc}
                                 Z^{(1)}_k \,e^{ i\Delta\omega(k)t-\frac{\Delta \Gamma(k)}{2}t} & 0 \\
                                  0& Z^{(2)}_k
\,e^{- i\Delta\omega(k)t+\frac{\Delta \Gamma(k)}{2}t} \\
                                \end{array}
                              \Bigg)~\mathcal{U}^{-1}\big(\theta_m(k)\big)~ \Psi^0_{\beta,\vec{k}}
                              \ee where \be \mathcal{U}\big(\theta_m(k)\big)= \Bigg(
                               \begin{array}{cc}
    \cos\theta_m(k) \big(1+i\widetilde{\gamma}(k)\big) & \sin\theta_m(k) \big(1-i\widetilde{\gamma}(k)\big) \\
                                                                                  - \sin\theta_m(k) & \cos\theta_m(k) \\
                                                                                \end{array}
                                                                              \Bigg)\label{Uma}
                               \ee
\be \mathcal{U}^{-1}\big(\theta_m(k)\big) =
\frac{1}{\big(1+i~\cos\theta_m(k)~ \widetilde{\gamma}(k)\big)}\Bigg(
                               \begin{array}{cc}
    \cos\theta_m(k)  & -\sin\theta_m(k) \big(1-i\widetilde{\gamma}(k)\big) \\
             - \sin\theta_m(k) & \cos\theta_m(k)\big(1+i\widetilde{\gamma}(k)\big) \\
                                                                                \end{array}
                                                                              \Bigg)\label{Umamin1}\ee

 Obviously the matrix $\mathcal{U}$ is \emph{not unitary}.

\subsection{Long time dynamics: Weisskopf-Wigner
Hamiltonian}

A phenomenological description of the dynamics of mixing and decay
for  neutral flavored mesons, for example $K_0
\overline{K}_0~;~B_0\overline{B}_0$  relies on the Weisskopf-Wigner
(WW) approximation\cite{WW1}. In this approximation the time
evolution of states is determined by a non-hermitian  Hamiltonian
that includes in a phenomenological manner the exponential
relaxation associated with the decay of the neutral mesons. This
approach has received revived  attention recently with the
possibility of observation of quantum mechanical coherence effects,
in particular CP-violation in current experiments with neutral
mesons\cite{WW2}. In ref.\cite{WW3} a field theoretical analysis of
the  (WW) approximation has been provided for the
$K_0\overline{K}_0$ system.

The form of the solution (\ref{posires2}) \emph{suggests} that a
(WW) approximate description of the asymptotic dynamics in terms of
a non-hermitian Hamiltonian is available.  To achieve this
formulation we separate explicitly the \emph{fast} time dependence
via the phase $e^{\mp i \overline{\omega}(k)t}$ for the positive and
negative frequency components, writing \be \Psi_{\vk}(t) = e^{- i
\overline{\omega}(k)t}~\Psi^+_{\vk}(t)+ e^{  i
\overline{\omega}(k)t}~\Psi^-_{\vk}(t) \label{slow} \ee where
$\Psi^{\pm}_{\vec k}(t)$ the amplitudes of the flavor vectors that
evolve \emph{slowly} in time. The solution for the positive
frequency component (\ref{posires}) follows from  the time evolution
of the \emph{slow} component determined by \be \label{sloweq} i
\frac{d}{dt} \Psi^+_{\vk}(t) = \mathcal{H}_{w}\Psi^+_{\vk}(t) \ee
where the effective \emph{non-hermitian} Hamiltonian $\mathcal{H}$
is given by \be \label{effH} \mathcal{H}_w= \frac{\delta
M^2}{4\,\overline{\omega}(k)} \Big(
\begin{array}{cc}
  -\cos2\theta & \sin2\theta \\
  \sin2\theta  & \cos2\theta \\
\end{array}
\Big) + \frac{\Sigma_R(k;\ovom)+\mathds{V}}{2\,\ovom}-i\,
\frac{\Sigma_I(k;\ovom)}{2\,\ovom }\ee with \bea
\Sigma_{R}(k;\ovom)+\mathds{V} & = &  \Bigg(
\begin{array}{cc}
 \Sigma_{R, ee}(k;\ovom)+V_{ee}& 0 \\
 0  & \Sigma_{R ,\mu\mu}(k;\ovom)+V_{\mu\mu} \\
\end{array}
\Bigg)\\\Sigma_{I}(k;\ovom)  & =      & \Bigg(
\begin{array}{cc}
 \Sigma_{I, ee}(k;\ovom) & 0 \\
 0  & \Sigma_{I ,\mu\mu}(k;\ovom)  \\
\end{array}
\Bigg) \eea

The (WW) Hamiltonian $\mathcal{H}_w$ can be written as \be
\mathcal{H}_w =   \Big\{
\frac{\overline{\Sigma}_R(k;\overline{\omega}(k))}{4\,\overline{\omega}(k)}
-i\frac{\overline{\Gamma}(k)}{2} \Big\} \mathds{1} + \frac{\delta
M^2 ~\overline{\varrho}(k)}{4~\ovom} \, \Bigg(
\begin{array}{cc}
  -(\cos2\theta_m(k)+i\widetilde{\gamma}(k)) & \sin2\theta_m(k) \\
  \sin2\theta_m(k) & (\cos2\theta_m(k)+i\widetilde{\gamma}(k)) \\
\end{array}\Bigg) \ee  where we have used the definitions given in
equations
(\ref{varrho},\ref{varrhok},\ref{tetamedav},\ref{gamatil},\ref{medGa}).
It is straightforward to confirm that the Wigner-Weisskopf
Hamiltonian can be written as \be \mathcal{H}_w =
\mathcal{U}(\theta_m(k)) ~ \Bigg(
\begin{array}{cc}
  \lambda_-(k)& 0 \\
 0 & \lambda_+(k) \\
\end{array}\Bigg) ~ \mathcal{U}^{-1}(\theta_m(k))\ee where $\mathcal{U}(\theta_m(k))$ is given in (\ref{Uma}) and
using the definitions given in eqn. (\ref{bigOm}-\ref{deltom})  the
\emph{complex} eigenvalues are \be  \label{lambdas} \lambda_\mp(k) =
W(k)-\ovom - i\frac{\overline{\Gamma}(k)}{2} \mp \left(\Delta
\omega(k) + i \frac{\Delta \Gamma(k)}{2}\right)   \ee

The solution of the effective equation for the slow amplitudes
(\ref{sloweq}) coincides with the long time dynamics given by
(\ref{posires}) when the wave function renormalization constants are
approximated as $Z^{(i)}_{\vk} \sim 1$. Therefore the (WW)
description of the time evolution based on the \emph{non-hermitian}
Hamiltonian $\mathcal{H}_w$ (\ref{effH}) \emph{effectively}
describes the evolution of flavor multiplets under the following
approximations:
\begin{itemize}
\item{Only the long-time dynamics can be extracted from the
 Weisskopf-Wigner Hamiltonian.}

\item{The validity of the perturbative expansion, \emph{and} of the
condition $\delta M^2 \ll \ovom^2$.}

\item{Wavefunction renormalization corrections are neglected
$Z^{(i)}\sim 1 $ and only leading order  corrections of order
$\widetilde{\gamma}(k)$ are included.}

\end{itemize}
 While the Weisskopf-Wigner effective description describes the
 relaxation of the flavor fields, it misses the stochastic noise
 from the bath, therefore, it \emph{does not} reliably describe the
 approach to equilibrium.

\section{Equilibration: effective Hamiltonian in the medium}\label{sec:equilibration}

As discussed in section \ref{equil} we study equilibration by
focusing on the asymptotic long time behavior of the one-body
density matrix or equal time correlation function, namely \be
\lim_{t\rightarrow \infty} \,  \langle \Psi_{\alpha,\vk}(t)
\Psi_{\beta,-\vk}(t) \rangle \,. \ee  In particular we seek to
understand    which basis diagonalizes   the equilibrium density
matrix.

Consider  general   initial conditions $\Psi^0 \neq 0$ and $\Pi^0
\neq 0$, in which case the flavor field $\Psi_{\alpha,\vk}(t)$ is
given by Eq. (\ref{inhosolution}) with  $G_{f}(k;t)$ given by eqn.
(\ref{Gdecay}). For $t \gg \Gamma^{-1}_{1,2}$, the first two
contributions to (\ref{inhosolution}) which depend on the initial
conditions fall-off exponentially as $e^{-\frac{\Gamma_{1,2}}{2}t}$
and \emph{only} the last term, the convolution with the noise,
survives at asymptotically long time, indicating that the
equilibrium state is insensitive to the initial conditions as it
must be.

To   leading order in the perturbative expansion in $G$, and in the
limit $\delta M^2/\overline{\omega}^{\,2}(k) \ll 1$, we can
approximate $\theta^{(1)}_m(k) \approx \theta^{(2)}_m(k)=
\theta_m(k)$, where the effective mixing angle in the medium
$\theta_m(k) $ is determined by the relations (\ref{tetamedav}).
Similarly we can approximate the wave function renormalization
constants as $Z^{(1)}(k)\approx Z^{(2)}(k) = Z(k)$ with

\be Z(k) =
\Bigg[1-\frac{1}{2\omega}\Big(\overline{\Sigma}^{'}_R(k;\omega) -
(-1)^i \frac{\delta M^2}{2} \cos2\theta_m(k)
\Delta^{'}_R(k;\omega)\Big)\Bigg]^{-1}_{\omega=\overline{\omega}(k)}
\ee

\noindent where the prime stands for derivative with respect to
$\omega$. Thus, $G_f(k;t)$ and $G_m(k;t)$ are related by \be
G_f(k;t) \approx Z(k) \, U(\theta_m)\, G_m(k;t)\,U^{-1}(\theta_m)
\ee  where $G_m(k;t)$ is given by \bea G_m(k;t)   = & &
\Bigg(\begin{array}{cc}
                  \frac{\sin (\Omega_1(k)t)}{\Omega_1(k)}~e^{-\frac{\Gamma_1(k)}{2}t}
& 0 \\
                  0 & \frac{\sin
(\Omega_2(k)t)}{\Omega_2(k)}~e^{-\frac{\Gamma_2(k)}{2}t}
                \end{array} \Bigg) + \nonumber \\ & & \frac{\widetilde{\gamma}(k)}{2}~\sin2\theta_m(k)
                \left[e^{-\frac{\Gamma_2(k)}{2}t}\frac{\cos(\Omega_2(k)t)}{\Omega_2(k)}-e^{-\frac{\Gamma_1(k)}{2}t}\frac{\cos(\Omega_1(k)t)}{\Omega_1(k)} \right]
                \left(
                         \begin{array}{cc}
                           0 & 1 \\
                           1 & 0 \\
                         \end{array}
                       \right)\,
                .\eea

It is useful to define the quantities $ h_{m}(t,\omega) $ and
$\widetilde{\xi}_{\beta, \vec k}(\omega)$ as follows

\be h_{m}(t,\omega)= \int_0^{t}  \, e^{-i\omega t'} \,G_{m}(k;t)
\,dt'   \label{hm}\ee and \be \xi_{\beta,\vk}(t-t') =
\int_{-\infty}^{+\infty} \,e^{i\omega (t-t')}\,
\widetilde{\xi}_{\beta,\vec k}(\omega)\, d\omega, \label{hm}\ee with
the noise average in the flavor basis given by \be  \langle
\langle\,
\widetilde{\xi}_{\rho,\vec{k}}(\omega)\,\widetilde{\xi}_{\sigma,-\vec{k}}(\omega')\,\rangle
\rangle  = \widetilde{{\mathcal
K}}_{\rho\sigma}(k;\omega)\delta(\omega+\omega')   =
\mathrm{Im}\widetilde{\Sigma}^R_{\rho
\sigma}(k;\omega)\coth\left(\frac{\beta
\omega}{2}\right)\delta(\omega+\omega')\, .\ee  We find convenient
to introduce \be \label{Km} \widetilde{{\mathcal K}}_{m}(k;\omega) =
U^{-1}(\theta_m)\,\widetilde{{\mathcal K}}(k;\omega)\,U(\theta_m)\,.
\ee

 The approach to equilibrium for   $t>>\Gamma^{-1}_{1,2}$ can be
established from the unequal time two-point correlation function,
given by

\be  \lim_{t,t'\rightarrow \infty} \, \langle \Psi_{\alpha,\vk}(t)
\Psi_{\beta,-\vk}(t') \rangle  = Z^2(k)  U(\theta_m)
\Bigg\{\int_{-\infty}^{+\infty}
\,d\omega~~e^{i\omega(t-t')}~~h_m(\infty,\omega)
\,\widetilde{{\mathcal K}}_m(k;\omega) \,h_m(\infty,-\omega)\Bigg\}
U^{-1}(\theta_m) \label{corrfun} \ee

\noindent where we have taken the upper limit $t\rightarrow \infty$
in (\ref{hm}). The fact that the correlation function becomes a
function of the \emph{time difference}, namely time translational
invariant, indicates that the density matrix commutes with the total
Hamiltonian in the long time limit. The one-body density matrix is
obtained from (\ref{corrfun}) in the coincidence limit $t=t'$.

Performing the integration over $ \omega$, we obtain after a lengthy
but straightforward calculation

\bea \lim_{t\rightarrow \infty} \, \langle \Psi_{\alpha,\vk}(t)
\Psi_{\beta,-\vk}(t) \rangle  = Z^2(k)\,
U(\theta_m)\,\left(%
\begin{array}{cc}
  \Lambda_{11}(k) & \Lambda_{12}(k) \\
  \Lambda_{21}(k) & \Lambda_{22}(k) \\
\end{array}%
\right)\,U^{-1}(\theta_m); \label{flavorcorrelation} \eea wherein
\bea \Lambda_{11}(k)=\frac{1}{2\,\Omega_1(k)}\coth\left(\frac{\beta
\Omega_1(k)}{2}\right)~;~~\Lambda_{22}(k)=\frac{1}{2\,\Omega_2(k)}\coth\left(\frac{\beta
\Omega_2(k)}{2}\right), \eea  and to the leading order of $\delta
M^2/\overline{\omega(k)}^{\,2}\ll 1$, we find
$\Lambda_{21}=\Lambda_{12}(\Omega_1 \rightarrow \Omega_2)$ where
\bea
\Lambda_{12}(k)&=&\frac{1}{2\,\Omega_1(k)}\coth\left(\frac{\beta
\Omega_1(k)}{2}\right) \sin2\theta_m(k)\;
\frac{\widetilde{\gamma}(k)\,
\eta(k)}{1+\left(\,\eta(k)\,\right)^2}~;~~
\eta(k)=\frac{2\Omega_1(k)\left(\,\Gamma_1(k)+\Gamma_2(k)\,\right)}{\delta
M^2 \overline{\varrho}(k)}. \eea

Since $\widetilde{\gamma}(k)\ll 1$, it is obvious that
$\Lambda_{12}(k)$ and $\Lambda_{21}(k)$ are perturbatively small
compared with $\Lambda_{11}(k)$ and $\Lambda_{22}(k)$, in either
case $\eta(k) \gg 1$ or $\eta(k) \ll 1$.   The asymptotic one-body
density matrix (\ref{equione}) then becomes \be \rho_{\alpha,\beta}(k;\infty)  = U(\theta_m)\,\left(%
\begin{array}{cc}
  \frac{Z}{2\,\Omega_1(k)}\coth\left(\frac{\beta
\Omega_1(k)}{2}\right) & \epsilon \\
  \epsilon & \frac{Z}{2\,\Omega_2(k)}\coth\left(\frac{\beta
\Omega_2(k)}{2}\right) \\
\end{array}%
\right)\,U^{-1}(\theta_m)~~;~~\epsilon \lesssim \mathcal{O}(G^2) \ee
where we neglected corrections of $\mathcal{O}(G^2)$ in the diagonal
matrix elements.

Neglecting the perturbative off-diagonal corrections, the one-body
density matrix \emph{commutes} with the   \emph{effective
Hamiltonian in the medium} which in the flavor basis is given by

\be H_{eff}(k) = \overline{\omega}(k) \Bigg(\begin{array}{cc}
                                        1 & 0 \\
                                        0 & 1
                                      \end{array}\Bigg)+ \frac{\delta
M^2}{4\,\overline{\omega}(k)} \Bigg(
\begin{array}{cc}
  -\cos2\theta & \sin2\theta \\
  \sin2\theta  & \cos2\theta \\
\end{array}
\Bigg)+ \frac{1}{2\overline{\omega}(k)}\,\Bigg(
\begin{array}{cc}
 \Sigma_{R, ee}(k;\ovom)+V_{ee}& 0 \\
 0  & \Sigma_{R ,\mu\mu}(k;\ovom)+V_{\mu\mu} \\
\end{array}
\Bigg) \label{Hmed} \ee this effective in-medium Hamiltonian can be
written in a more illuminating
  form

  \be H_{eff}(k) =  U(\theta_m)\,\Bigg(
                                         \begin{array}{cc}
                                           \Omega_1(k) & 0 \\
                                           0 & \Omega_2(k) \\
                                         \end{array}
                                       \Bigg)\,U^{-1}(\theta_m)
                                       \label{Heffdiag}\ee where
                                       $\Omega_{1,2}(k)$ are the
 correct propagation frequencies in the medium given by eqn.
 (\ref{omegone},\ref{omegtwo}).

This effective Hamiltonian includes the radiative corrections in the
medium via the flavor diagonal self-energies (forward scattering)
and apart from the term proportional to the identity is identified
with the \emph{real part} of the  Weisskopf-Wigner Hamiltonian
$\mathcal{H}_w$ given by eqn. (\ref{effH}). This form
  highlights that the off-diagonal elements of the one-body density matrix in the basis
  of  \emph{eigenstates of the effective Hamiltonian in the medium} are
\emph{perturbatively small}. The unitary transformation
$U(\theta_m)$ relates the flavor fields to the fields in the
\emph{basis of the effective Hamiltonian in the medium}.

 Comparing this
result to the free field case in thermal equilibrium, where the one
body density matrix in the flavor basis is given by eqn.
(\ref{rho1flav}), it becomes clear that in the long time limit
equilibration is achieved and the one-body density matrix is nearly
diagonal in the basis of the \emph{eigenstates of the effective
Hamiltonian in the medium} (\ref{Hmed}) with the diagonal elements
determined by the distribution function of  these eigenstates.

This means that within the realm of validity of perturbation theory,
the equilibrium correlation function   \emph{is nearly diagonal in
 the basis of the effective Hamiltonian in the medium}.   This result confirms the
 arguments  advanced in
\cite{chargedlepton}. Since the effective action is quadratic in the
``neutrino fields'' higher correlation functions are obtained as
Wick contractions of the two point correlators, hence the fact that
the two point correlation function and consequently the one-body
density matrix are diagonal in the basis of the eigenstates of the
effective Hamiltonian in the medium guarantee that all higher
correlation functions are also diagonal in this basis.

\subsection{On ``sterile neutrinos''}\label{sec:sterile}

The results obtained in the previous sections apply to the case of
two ``flavored neutrinos'' both in interaction with the bath.
However, these results can be simply extrapolated to the case of one
``active'' and one ``sterile'' neutrino that mix via a mass matrix
that is off-diagonal in the flavor basis. By definition a
``sterile'' neutrino \emph{does not} interact with hadrons, quarks
or charged leptons, therefore for this species there are no
radiative corrections. Consider for example that the ``muon
neutrino'' represented by $\phi_\mu$ does \emph{not} couple to the
bath, but it does couple to the ``electron neutrino'' solely through
the mixing in the mass matrix. Since the interaction is diagonal in
the flavor basis, the decoupling of this  ``sterile neutrino'' can
be accounted for simply by imposing the following ``sterility
conditions'' for the matter potential $\mathds{V}$ and the self
energies

\be V_{\mu\mu} \equiv 0 ~;~\Sigma_{R,\mu\mu} \equiv
0~;~\Sigma_{I,\mu\mu}\equiv 0 \;. \label{sterility} \ee All of the
results obtained above for the dispersion relations and relaxation
rates apply to this case by simply imposing these ``sterility
conditions''. In particular it follows that \be
\Gamma_1(k)=\Gamma_{ee}(k)\cos^2\theta_m(k)~;~\Gamma_2(k)=\Gamma_{ee}\sin^2\theta_m(k)\label{sterigamas}
\ee \noindent where $\Gamma_{ee}(k)$ is the relaxation rate of the
\emph{active neutrino} in absence of mixing. This result highlights
that in the limit $\theta \rightarrow 0$   the in-medium eigenstate
labeled ``2'' is seen to correspond to the sterile state, because in
the absence of mixing this state does not acquire a width. However,
for non-vanishing vacuum mixing angle, the ``sterile neutrino''
nonetheless \emph{equilibrates} with the bath as a consequence of
the ``active-sterile'' mixing, which effectively induces a coupling
between the ``sterile'' and the
bath\cite{dm,aba,foot,raffsigl,dibari}. The result for
$\Gamma_2(k)$, namely the relaxation rate of the ``sterile''
neutrino is of the same form as that proposed in refs.
\cite{dm,aba,foot,raffsigl,dibari}. The result for the ``sterile''
rate $\Gamma_2(k)$ compares to those in these references in the
limit in which perturbation theory is valid, namely
$\Sigma_{ee}(k)/\delta M^2 \overline{\varrho}(k) \ll 1 $ since the
denominator in this ratio is proportional to the oscillation
frequency in the medium.

\section{Summary of results and conclusions}\label{sec:conclu}

In this article we studied the non-equilibrium dynamics of mixing,
oscillations and equilibration in a model field theory that bears
all of the relevant features of the standard model of neutrinos
augmented by a mass matrix off diagonal in the flavor basis. To
avoid the     complications associated with the spinor nature of the
neutrino fields, we studied an interacting  model of flavored
neutral mesons. Two species of flavored neutral mesons play the role
of two flavors of neutrinos, these are coupled to other mesons which
play the role of hadrons or quarks and charged leptons, via flavor
diagonal interactions that model charged currents in the standard
model. These latter meson fields are taken to describe a bath in
thermal equilibrium, and the meson-neutrino fields are taken to be
the ``system''. We obtain a reduced density matrix  and the
non-equilibrium effective action for the ``neutrinos''   by
integrating out the bath degrees of freedom up to second order in
the coupling in the full time-evolved density matrix.

The non-equilibrium effective action yields all the information on
the particle and quasiparticle modes in the medium, and the approach
to equilibrium.

{\bf Summary of results:}
\begin{itemize}
\item{  We obtain the dispersion relations, mixing
angles and relaxation rates of the two quasiparticle modes in the
medium. The dispersion relations and mixing angles are of the same
form as those obtained for neutrinos in a
medium\cite{notzold,boyho}.}

\item{The relaxation rates are found to be \bea \Gamma_1(k)  & = &   \Gamma_{ee}(k)
\cos^2\theta_m(k)+\Gamma_{\mu\mu}(k)\sin^2\theta_m(k)
\label{Gama1ee} \\\Gamma_2(k)  & = &   \Gamma_{\mu\mu}(k)
\cos^2\theta_m(k)+\Gamma_{ee}(k)\sin^2\theta_m(k)
\label{Gama2mumu}\eea where \be\Gamma_{\alpha\alpha}(k) =
\frac{\Sigma_{I,\alpha\alpha}(k;\ovom)}{ \ovom}\ee are the
relaxation rates of the flavor fields in \emph{absence} of mixing
and $\Sigma_{I,\alpha,\alpha}$ are the imaginary parts of the
``neutrino'' self energy which is diagonal in the flavor basis.
These relaxation rates are similar in form to those proposed  in
refs.\cite{raffsigl,aba,foot,dibari}, within the context of
active-sterile conversion or flavor conversion in supernovae.}

\item{The long time dynamics is approximately described by an
effective Weisskopf-Wigner approximation with a non-hermitian
Hamiltonian. The real part includes the ``index of refraction'' and
the renormalization of the frequencies and the imaginary part is
determined by the absorptive part of the second order self-energy
and describes the relaxation. While this (WW) approximation
describes mixing, oscillations and relaxation, it \emph{does not}
capture the dynamics of equilibration.}

\item{ For time $t>> \Gamma^{-1}_{1,2}$ the two point function of
the neutrino fields becomes time translational invariant reflecting
the approach to equilibrium. The asymptotic long time limit of the
one-body density matrix reveals that the density matrix is nearly
diagonal in the    basis of eigenstates of an effective Hamiltonian
in the medium (\ref{Hmed}) with perturbatively small off-diagonal
corrections in this basis. The diagonal components in this basis are
determined by \emph{the distribution function of these
  eigenstates}. }

\item{  ``Sterile'' neutrinos: these results   apply to the case in which only one of the flavored neutrinos
is ``active'' but the other is ``sterile''. Consider for example
that the ``muon neutrino'' is sterile in the sense that it does not
couple to the bath. This sterile degree of freedom is thus
identified with the in medium eigenstate ``2'' because in the
absence of mixing  $\theta=0$ its dynamics is completely free.  The
``sterility'' condition corresponds to setting the matter potential
$V_{\mu\mu}=0$ and the self-energy $\Sigma_{\mu\mu}=0$ with a
concomitant change in the dispersion relations. All the results
obtained above apply just the same, but with $\Gamma_{\mu\mu}(k)=0$,
from which it follows that
$\Gamma_2(k)=\Gamma_{ee}(k)\sin^2\theta_m(k)$. The final result is
that ``sterile'' neutrinos do thermalize with the bath via
``active-sterile'' mixing. If the mixing angle in the medium is
small, the equilibration time scale for the ``sterile neutrino'' is
much larger than that for the ``active'' species, but equilibration
is eventually achieved nonetheless. This result is a consequence of
``active-sterile'' oscillations which effectively induces an
interaction of the sterile neutrino with the
bath\cite{dm,aba,foot,raffsigl,dibari}. }

\end{itemize}

Although the meson field theory studied here describes quite
generally the main features of mixing, oscillations and relaxation
of neutrinos, a detailed quantitative assessment of the relaxation
rates and dispersion relations do  require a full calculation in the
standard model. Furthermore there are several aspects of neutrino
physics that are distinctly associated with their spinorial nature
and cannot be inferred from this model. While only the left handed
component of neutrinos couple to the weak interactions, a (Dirac)
mass  term couples the left to the right handed component, and
through this coupling the right handed component  develops a
dynamical evolution. Although the coupling to the right handed
component is very small in the ultrarelativistic limit, it is
conceivable that non-equilibrium dynamics may lead to a substantial
right handed component during long time intervals. The study of this
possibility would be of importance in the early Universe because the
right handed component may thereby become an ``active'' one that may
contribute to the total number of species in equilibrium in the
thermal bath thus possibly affecting the expansion history of the
Universe.


Another important fermionic aspect is Pauli blocking which is
relevant in the case in which neutrinos are degenerate, for example
in supernovae.


These aspects will be studied elsewhere.

\begin{acknowledgments}   The authors acknowledge support from the U.S. NSF
under Grants No. PHY-0242134 and No. 0553418. C. M. Ho acknowledges
partial support through the Andrew Mellon Foundation and the
Zaccheus Daniel Fellowship.
\end{acknowledgments}

\end{document}